\documentclass{article}

\usepackage[utf8]{inputenc}
\usepackage{amsmath}
\usepackage{amsthm}
\usepackage{amssymb}
\usepackage{graphicx}
\usepackage{physics}
\usepackage{todonotes}
\usepackage{pgf,tikz,pgfplots}
\usepackage[english]{babel}
\usepackage[margin=1.1in]{geometry}
\usepackage{authblk}
\pgfplotsset{compat=1.16}
\usepackage{tabularx}
\usepackage{siunitx}

\newcounter{protocol}
\newenvironment{protocol}[1]
  {\par\addvspace{\topsep}
   \noindent
   \tabularx{\linewidth}{@{} X @{}}
    \hline
    \refstepcounter{protocol}\textbf{Protocol \theprotocol} #1 \\
    \hline}
  {\\
  \hline
   \endtabularx
   \par\addvspace{\topsep}}

\begin{document}
\title{Quantum City: 
simulation of a practical near-term metropolitan quantum network}
\author[1,2]{Raja Yehia}
\author[1]{Simon Neves}
\author[1]{Eleni Diamanti}
\author[2]{Iordanis Kerenidis}
\affil[1]{Sorbonne Université, CNRS, LIP6, F-75005 Paris, France}
\affil[2]{Université Paris Cité, CNRS, IRIF, F-75013 Paris, France}
\date{\today}

\maketitle

\begin{abstract}
We present the architecture and analyze the applications of a metropolitan-scale quantum network that requires only limited hardware resources for end users. Using NetSquid, a quantum network simulation tool based on discrete events, we assess the performance of several quantum network protocols involving two or more users in various configurations in terms of topology, hardware and trust choices. Our analysis takes losses and errors into account and considers realistic parameters corresponding to present or near-term technology. Our results show that practical quantum-enhanced network functionalities are within reach today and can prepare the ground for further applications when more advanced technology becomes available.
\end{abstract}

\section{Introduction}

Quantum communication networks promise to make a wide range of applications available to parties that can generate, share, store, and manipulate quantum information. Such networks can be seen as infrastructures that progressively advance in terms of the quantum technology they integrate and, consequently, of the  functionalities and user applications they enable~\cite{QIavision}. Photonic networks targeting specific applications like quantum key distribution (QKD) are already under deployment today~\cite{ChinaQKDNetwork,OpenQKD,BristolQCity}, while the Quantum Internet in its full-fledged vision~\cite{QIavision, QIRG} will still require some years to materialize.\\

A large-scale quantum information network, where entanglement can efficiently be routed between all pairs of users over long distances in a modular way, will open the way to the demonstration of numerous protocols featuring a security, efficiency or computational quantum advantage. These include, for instance, conference key agreement \cite{Murta_2020}, distributed consensus \cite{Consensus} or ranking \cite{Ranking}, blind and verifiable delegated computing \cite{DelegatedQC}, clock synchronisation \cite{clocksync}, anonymous transmission \cite{Anonymity}, quantum money \cite{QuantumMoney}, quantum e-voting \cite{fedeVoting}, distributing sensing \cite{DistributedSensing} and computing, and many more. The required technologies for the realization of such a network are the subject of intense research efforts worldwide, with promising results in high-efficiency quantum memories and elementary quantum repeaters and networks~\cite{OpticaLKB,ICFO,Qutech}. However, important challenges still need to be overcome until a useful quantum advantage for advanced tasks can be shown in practice using quantum repeater links~\cite{repeaterNV, TimReapeaterChain}.\\



In this work, we wish to show that several quantum network applications are in fact accessible even with near-term technology. Our goal is to contribute to the identification of network topologies and system architectures that can enhance today's communications with quantum-enabled functionalities in a realistic and practical way, while more advanced technologies gradually become available and upgrade the network.
Assuming, for example, that all nodes of the network possess the capabilities provided by a full quantum computer, or even that each user accessing the network is able to control as many qubits as necessary for any application and has access to unlimited quantum storage time, is presently unrealistic but also not necessary for many applications. It is widely accepted that at least for the foreseeable future, quantum computers will be accessible remotely at a few locations, and protocols like blind delegation will allow a user with limited quantum hardware to enjoy computation capacities of a distant quantum processor without revealing the nature or result of the computation. Moreover, many communication protocols, as we will see later on, can be performed by users with relatively simple quantum capabilities, which could be available as compact, portable devices in the near future.\\

Starting from this premise, we propose here the Quantum City as a metropolitan-scale quantum communication network with optimized resources that is both realizable with current or near-term quantum technology and sufficiently advanced to support several interesting quantum network applications. In this work, we define the main ingredients of such a network, provide examples of supported protocols, perform extensive network simulations and also show how this architecture can fit in the framework of the full-scale Quantum Internet. Our analysis uses as a running example a quantum network in the Paris region, featuring characteristics common to many urban areas in Europe and beyond.\\

In general terms, the Quantum City consists of two types of nodes: the \textbf{Qonnector}, a powerful network node that has the ability to generate and share bi or multi-partite entanglement to users through quantum channels; and the \textbf{Qlients}, network users with limited capabilities, who can generate, manipulate, and measure single photons. At metropolitan scale, the network will consist of a single or a few Qonnectors, using optical fibers to connect to a few hundred or thousand Qlients. The underlying principle is that the topology and overall architecture is flexible enough to allow for the implementation of various protocols enabling different functionalities and corresponding applications.\\


In the following sections, we first describe in more detail the capabilities of the Qonnector and Qlient nodes, and also recall the main properties of some protocols that can be supported by these capabilities. Then, we use the discrete-event quantum network simulation tool NetSquid~\cite{coopmans2021netsquid,Netsquid} to assess the feasibility and performance of selected protocols over our baseline Paris region network, assuming Qlients equipped with state-of-the-art photonic technology. Our benchmark quantities include the sifted rate, throughput and quantum bit error rate (QBER) that can be obtained with realistic physical parameters and quantum network models. Our results provide evidence that metropolitan-scale quantum network realizations with resources available today or in the near future can already have significant impact, offering concrete quantum-enabled functionalities and applications. Crucially, we believe that this architecture allows  for growth and adaptability to future developments, as well as for hardware heterogeneity.\\

In a companion paper that will appear shortly, we also study how Quantum Cities can be connected together via satellite links using as a running example a Europe-wide quantum communication network, a Quantum Europe. 



\section{The Quantum City architecture}

\subsection{Description}
\begin{figure}[!ht]
    \centering
    \includegraphics[width = 9.5cm]{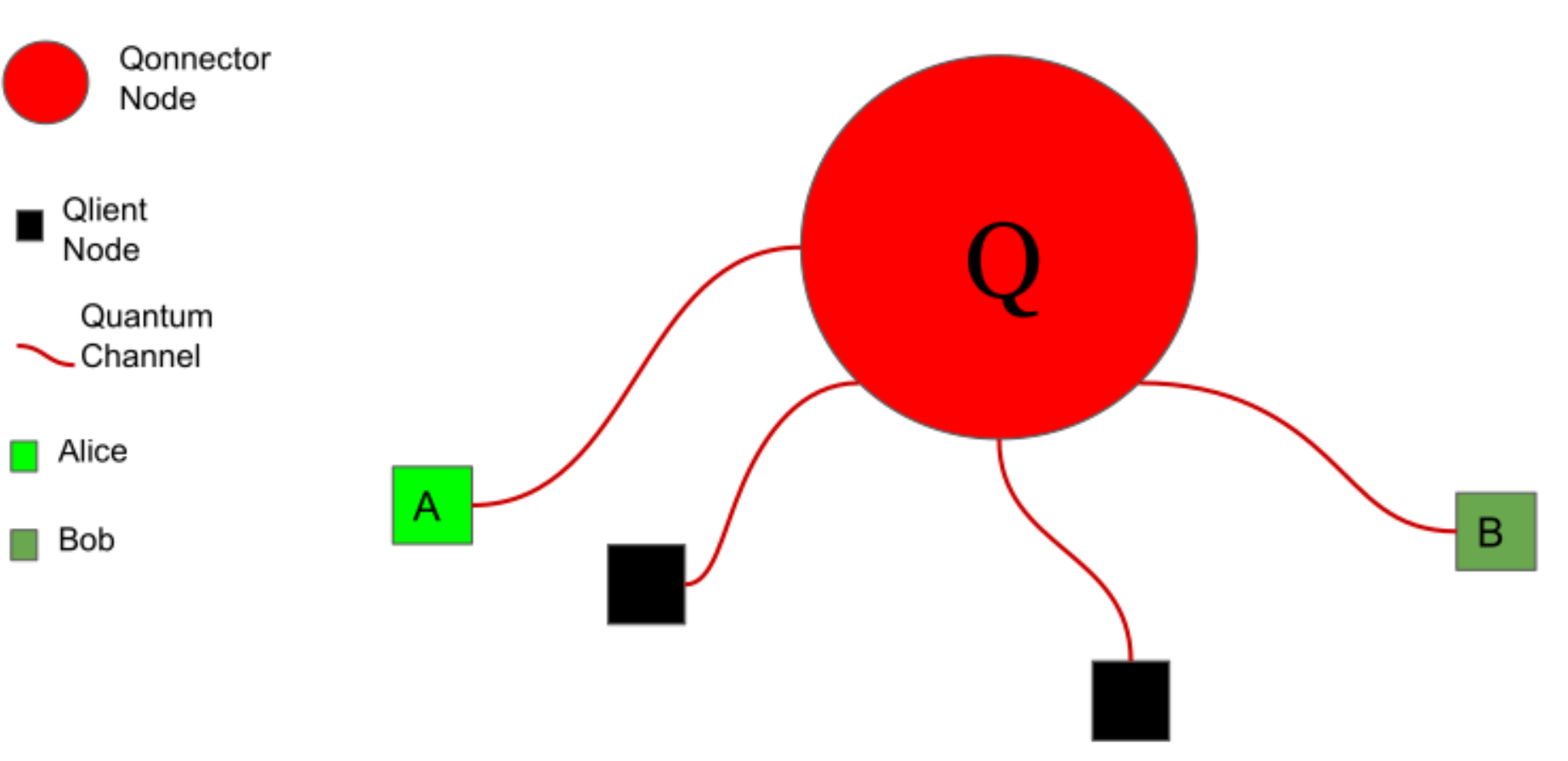}
    \caption{The Quantum City topology. It is a star-type quantum network with a special node at its centre, the Qonnector, which has advanced quantum capabilities. Each end node (Qlient, e.g. Alice and Bob) has limited quantum capabilities and is connected to the Qonnector through an optical fiber. The Quantum City allows for Qlients to perform quantum network protocols via the Qonnector.}
    \label{fig:ArQitecture}
\end{figure}

In its basic form, the Quantum City has a star topology with a central node that we call Qonnector linked to a number of users that we call Qlients through optical fibers (see Fig.~\ref{fig:ArQitecture}). This allows for centralized routing procedures and asymmetric distribution of hardware between a powerful Qonnector and more limited Qlients. This corresponds well to the expected intermediate-term development of quantum networks, where some nodes with advanced quantum resources will be providing quantum access and functionalities to a number of users with limited quantum capabilities. Below we describe precisely the abilities of a Qonnector and a Qlient node in our model.\\ \newpage

\textbf{Qonnector} nodes provide the core quantum functionalities of our model. They are an abstraction of quantum servers providing quantum services to end users. The capabilities of a Qonnector may vary and evolve in time, but in our analysis they are nodes that can create and share photonic entangled quantum states and are connected both classically and quantumly to a certain number of users from whom they can receive photons and perform measurements on them. Using state-of-the-art technology, we can already assume that a Qonnector has the following capabilities: to generate and manipulate single-qubit states as well as entangled states such as Bell pairs and few-qubit GHZ states; 
to receive and measure photonic states and to perform probabilistic Bell state measurements on two photons arriving simultaneously; 
to route photonic states arriving from one Qlient to another Qlient.
A Qonnector will also use classical computing power and classical Internet. In particular, a Qonnector holds a list of each Qlient's identification and port to use to communicate with each of them. It also has classical memory slots reserved for each connected Qlient to perform classical pre and post-processing in protocols. This centralizes network information in one node, facilitating routing of quantum information and addition of new Qlients to the network. In a metropolitan area, a Qonnector able to perform those specific tasks provides the resource for the creation of a quantum network of tens or hundreds of users. A Qonnector thus represents a network provider for an area.

As mentioned before, one can imagine a more powerful Qonnector, for example a node equipped with quantum memories to enable efficient, on-demand operations, or with a quantum processor to which a Qlient can delegate securely a computation. Distant Qonnectors may also be linked with quantum repeater or satellite links, forming a backbone network with entangled central nodes. Our architecture is agile enough to handle such upgrades in the quantum capabilities of the Qonnector nodes, while making it possible for the quantum network to support a number of different functionalities already with a simpler node as a Qonnector and Qlients with limited, realistic quantum capabilities.\\

\textbf{Qlient} nodes represent the end users connecting to the quantum network. They abstract private users that hold quantum communication devices that can be commercially available in the near future for wide use. They are classically connected to the rest of the network through the classical Internet and have usual classical computing power. We assume that they have limited quantum hardware capabilities, namely that they can manipulate a single qubit at a time. More precisely, they can generate, receive, and measure any single-qubit photonic state (in fact, it may be sufficient to have the ability to either generate or receive and measure such states). In a more advanced version, Qlients also have the capability to store quantum states for a short period of time. Industrial-grade devices offering these capabilities are already available today or will become in the near future, and can be expected to become more suitable for wider use in the following years, thanks to advances for instance in photonic integration~\cite{PhotonicIntegrationRoadmap}.
As we will show in the following, Qlients with these capabilities can already have access to various quantum network applications.



\subsection{Modelling quantum network processes}
\label{sec:model}

\begin{figure}[!ht]
    \centering
    \includegraphics[width= 13cm]{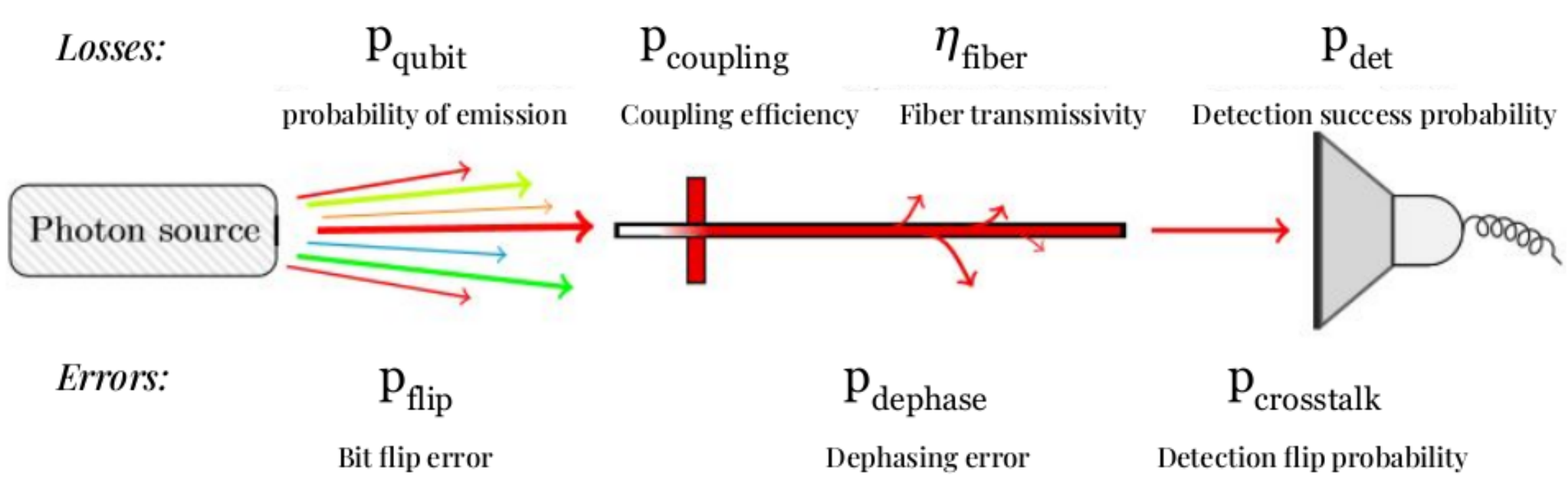}
    \caption{Photon loss and error model for the emission, transmission and measurement of a qubit.}
    \label{fig:lossmodel}
\end{figure}

Let us now go into more detail in the operations that the nodes of our network architecture can perform, to include in particular losses and errors that are inherent in any realistic quantum operation.
In this work, we model losses and errors through depolarizing and dephasing channels that act on the state $\rho$ on which the operation is applied: 
\begin{equation}
\mathcal{D}^{\lambda_{1}}_{\textrm{depol}}(\rho) = \lambda_{1}\rho + (1-\lambda_{1})\frac{\mathbb{I}}{d}\ ,
\end{equation}
\begin{equation}
\mathcal{D}^{\lambda_{2}}_{\textrm{dephase}}(\rho) = \lambda_{2}\rho + (1-\lambda_{2})Z\rho Z\ ,
\end{equation}
where $\lambda_1$ and $\lambda_2$ represent losses and noise, respectively. Every time a specific operation is applied to a qubit, these channels describe the applied operation with its corresponding parameters. In other words, $\lambda_1$ corresponds to a loss probability and $\lambda_2$ to an error probability. 
More precisely, we consider the following sources of losses and errors:

\begin{itemize}
    \item The creation of any single-qubit state $\ket{\psi}=\cos\theta\ket{0} +\sin\theta\ket{1}$ is attempted at a rate $f_{\text{qubit}}$ and succeeds with probability $p_{\text{qubit}}$. A bit flip error occurs with probability $p_{\text{flip}}$.
    \item The creation of an EPR pair is attempted at a rate $f_{\text{EPR}}$ and succeeds with probability $p_{\text{EPR}}$.
    \item The creation of an $n$-qubit GHZ state is attempted at a rate $f_{\text{GHZ}-n}$ and succeeds with probability $p_{\text{GHZ}-n}$.
    \item The routing of a state received from a user to another one succeeds with probability $p_{\text{transmit}}$.
    \item A Bell State Measurement (BSM) on two photonic states received simultaneously succeeds with probability $p_{\text{BSM}}$.
    \item Photonic qubits are coupled in fibers with a probability $p_{\text{coupling}}$, also called coupling efficiency.
    \item Losses in optical fibers are characterized by the quantity $\eta_{\text{fiber}}$ in $\si{\dB/\kilo\meter}$, such that a photon is transmitted over a distance $L$ with probability $\exp(-\eta_{\text{fiber}}\cdot L/10)$, and dephasing occurs with probability $p_{\text{dephase}}$. 
    \item A photonic qubit is successfully measured with probability $p_{\text{det}}$, and the outcome is flipped with probability $p_{\text{crosstalk}}$.
    \item Detectors are active only in a time window $\Delta t_{\text{det}}$ around each state creation attempt, called the detection gate.
    \item Detectors can spontaneously be triggered even in the absence of photons, resulting in dark counts at an average rate $R_{\text{dark}}$. Hence, they occur with a probability $p_{\text{dark}} = R_{\text{dark}} \cdot \Delta t_{\text{det}}$ when attempting to create a state. Dark counts typically trigger state detection, when none was emitted or when the state was lost. They can also lead to double outcomes at the detection of one qubit, in which case the data is discarded.
\end{itemize}
Other effects can occur with far lower probabilities and are therefore ignored. In Fig.~\ref{fig:lossmodel}, we show as an example the error and loss model that we consider for sending and receiving a photonic qubit. \\

For modularity, all the failure probabilities and rates can be defined separately for each Qlient and for the Qonnector. This reflects the realistic case of a difference in hardware performance between a server node and the various end users. 
Note that we ignore here the travel time of photons. We fix it to 1~ns in our simulations, regardless of the distance between the nodes. In a real life setting, this travel time should be carefully measured so that receiving nodes know when they should expect photons. This would typically cause the receiving node to start its measurement routine a few milliseconds after the sending node starts sending photons, to account for photon travel time in the fiber.

To simulate our architecture we use a quantum network simulation tool using discrete events, called NetSquid \cite{coopmans2021netsquid}. Most errors and losses are embedded in NetSquid, which facilitates the simulation of realistic quantum networks. The modularity of this simulator and its built-in components allows us to easily create a network model on which one can run protocols of interest. By defining simple local routines for each node such as the creation and sending of a state or performing a Bell state measurement, we can simulate the quantum operations involved in the protocols we consider. In this work, we focus on modeling quantum operations and we do not take classical pre and post-processing into account assuming that they can be performed fast enough by any classical computer. This will allow us to find the critical physical parameters of the Quantum City architecture that limit the functionalities accessible to a user, hence pointing to possible optimizations for networks under development. To learn more about the use of NetSquid, we encourage the reader to consult the NetSquid website \cite{Netsquid}. The code used in our work is available on GitHub \cite{github}.

\section{Network protocols in a Quantum City}

The Quantum City architecture is designed to be modular: for one functionality the Qlients can choose between different network protocols depending on their hardware or on the level of trust they want to put into their Qonnector. This is crucial to allow our architecture to grow with the research and adapt to new protocols that will be developed. 
In this section we go over some of the protocols that are available to Qlients in our setting. We will simulate them in the next section. 

\subsection{Quantum Key Distribution}
\label{sec:QKD}

\begin{figure}[!ht]
    \centering
    \includegraphics[width = 15cm]{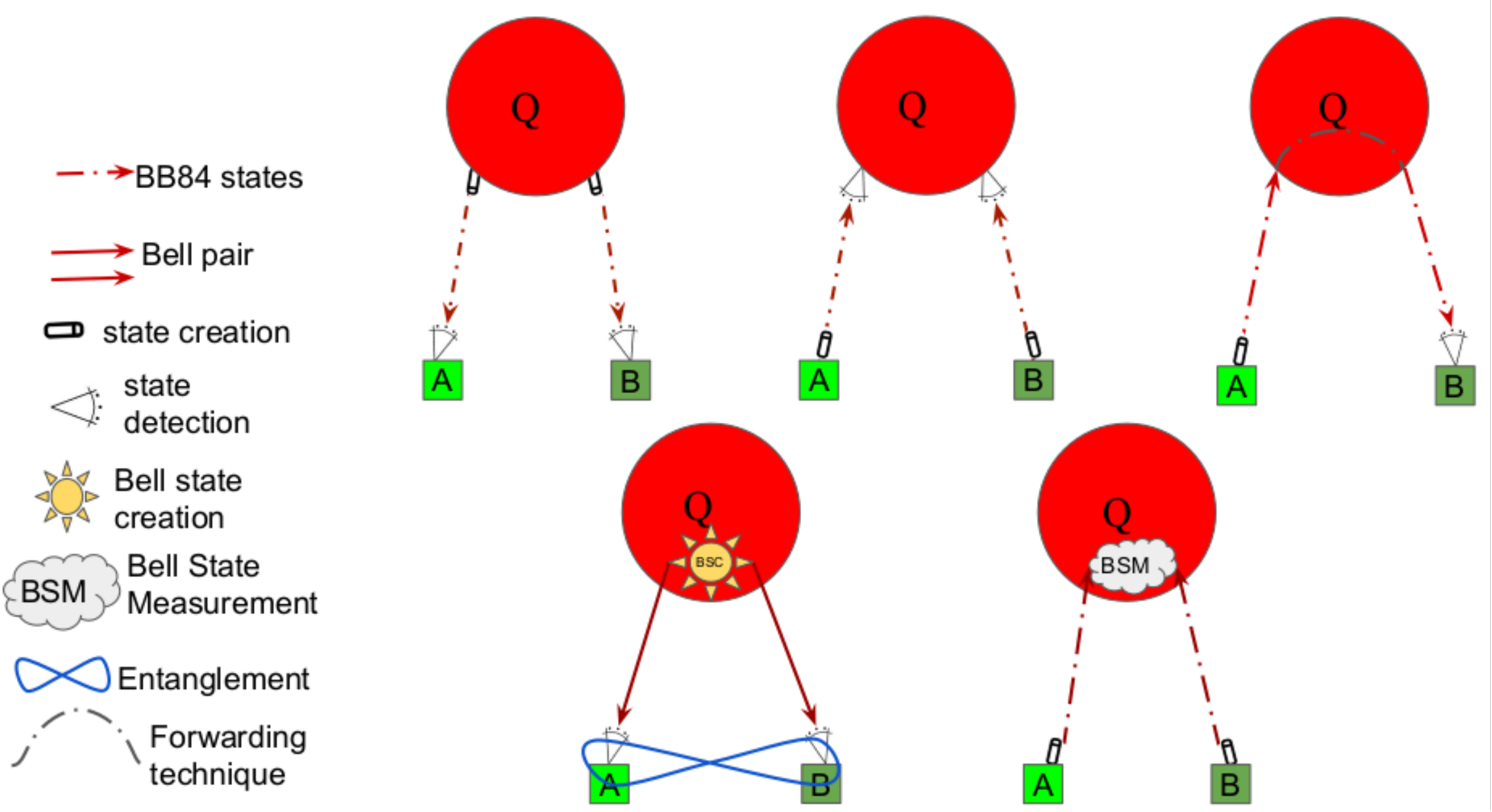}
    \caption{From left to right, top to bottom: 2xBB84, 2xBB84 reversed, transmitted BB84, entanglement-based QKD and MDI-QKD.}
    \label{fig:AllQKD}
\end{figure}
Quantum key distribution (QKD) is a fundamental quantum network application, which allows two parties to share a secret key with information-theoretic security~\cite{SecurityQKD}. This key can be subsequently used for applications such as secure communication or long-term secure storage. QKD protocols are well understood and have been studied extensively~\cite{DLQY:npjQI16,pir2020}, and as a result the achieved key rate for various protocols is an excellent benchmark for quantum networks.\\

In our analysis, we will simulate for our Quantum City network five protocols implementing the QKD functionality (see Fig.~\ref{fig:AllQKD}). These protocols differ in their trust assumptions, hardware requirements and key rate performance. We stress here again the importance of a modular network supporting different protocols and allowing for resource optimization depending on the user choices and hardware upgrades. We will focus on the setting where two Qlients, Alice and Bob, are connected to a Qonnector and wish to share a secret key. Our simulation will allow us to play with different parameters such as the distance between the nodes or the error or rate of each operation described in Sec.~\ref{sec:model}.



\subsubsection{BB84}

The most standard way for two Qlients, Alice and Bob, to share a secret key is to perform separately the BB84 protocol \cite{BENNETT20147} with the Qonnector. The Qonnector can then transmit, for instance, Alice's key to Bob as a secret message using Bob's key. This requires that Alice and Bob have complete trust over the Qonnector. 
We recall the quantum transmission part of the BB84 protocol between a Qonnector and a Qlient in our setting when the Qonnector has the role of the sender and the Qlient is the receiver.
\begin{protocol}{BB84}

\begin{enumerate}
    \item At each timestep defined by $1/f_{\text{qubit}}$, the Qonnector chooses a random bit $b\in\{0,1\}$ and a random basis $t \in \{\cross,+\}$ and creates qubit $\ket{0}$ if $b=0$ and $t=\cross$, $\ket{1}$ if $b=1$ and $t=\cross$, $\ket{+}$ if $b=0$ and $t=+$ and $\ket{-}$ if $b=1$ and $t=+$. 
    \item The Qonnector sends the state to the Qlient through the quantum channel.
    \item At each timestep, the Qlient randomly chooses a basis in $\{\cross,+\}$ and performs a measurement on her qubit. 
    Whenever she gets a click, she records the outcome as well as the current timestep.
    \end{enumerate}

\end{protocol}
After a fixed number of timesteps, the Qonnector and the Qlient perform the necessary classical post-processing steps, namely sifting, parameter estimation, error correction and privacy amplification, using authenticated classical communication.
Upon completion of this protocol with both Alice and Bob, and the actions of the Qonnector as a trusted node, the two users share a secret key. \\

Note that the role of sender and receiver can be switched between the Qlients and the Qonnector (see Fig.~\ref{fig:AllQKD}, 2xBB84 reversed). This does not affect the performance but can be interesting for removing the need for Qlients to hold single-photon detection technology, which can be a demanding hardware condition presently.

\subsubsection{Transmitted BB84}

Another configuration based on the same protocol that bypasses the necessity for a trusted node is to set one Qlient as the sender and the other as the receiver. The Qonnector then serves as a simple relay, transmitting the qubit received from one Qlient to the other. The description of the quantum transmission part for this protocol configuration is as follows:

\begin{protocol}{Transmitted BB84}
\begin{enumerate}
    \item At each timestep defined by $1/f_{\text{qubit}}$, the sending Qlient randomly chooses a random bit $b\in\{0,1\}$ and a random basis $t \in \{\cross,+\}$ and creates qubit $\ket{0}$ if $b=0$ and $t=\cross$, $\ket{1}$ if $b=1$ and $t=\cross$, $\ket{+}$ if $b=0$ and $t=+$ and $\ket{-}$ if $b=1$ and $t=+$. He sends it to the Qonnector.
    \item The Qonnector links its receiving ports for the sending Qlient and its sending port for the receiving Qlient. The qubit is thus transmitted to the receiving Qlient.
    \item At each timestep, the receiving Qlient randomly chooses a basis in $\{\cross,+\}$ and performs measurement on her qubit. Whenever she gets a click, she records the outcome as well as the current timestep.
\end{enumerate}

\end{protocol}

The full QKD protocol in this case is performed directly between the two Qlients. The Qonnector here is part of the quantum channel and its actions, including any potential tampering, are taken into account in the security analysis. 
In this relay setting, the key rate will be lower as the distance travelled by the qubits between the sending and receiving Qlients via the Qonnector is longer.

\subsubsection{Entanglement-based QKD}

In entanglement-based QKD \cite{Ekert,BBM92}, the Qonnector node does not need to be trusted. It sends EPR pairs to Alice and Bob, who then measure the qubits they receive in a randomly selected basis depending on the protocol. The main difficulty comes from the fact that both photons from the pair should arrive at their destination. For the BBM92 protocol and assuming the Bell state $\ket{\psi^-}$ shared between the users, we have:

\begin{protocol}{BBM92}
\begin{enumerate}
    \item At each timestep defined by $1/f_{\text{EPR}}$, the Qonnector prepares an EPR pair in the $\ket{\psi^-}=\frac{1}{\sqrt{2}}(\ket{01}-\ket{10})$ state.
    \item The Qonnector sends one photon of the pair to Alice and the other to Bob.
    \item Both Alice and Bob receive and measure each qubit randomly in the $\{\cross,+\}$ basis. If they get a click, they record the output and the timestep.
\end{enumerate}

\end{protocol}
After performing this protocol for some predefined time, Alice and Bob proceed with the classical post-processing, which assumes again the existence of authenticated classical communication. 
Entanglement-based QKD is an attractive candidate for network deployment, as high-performance photonic Bell state generation is now a readily available technology and, as already mentioned, the intermediate node does not need to be trusted. In this setting, Qlients need to have access to single-photon detection technology.

\newpage

\subsubsection{Measurement-Device-Independent QKD}

Measurement-Device-Independent (MDI) QKD \cite{MDIQKD,Pirandola2} allows Alice and Bob to share a secret key by sending qubits to their Qonnector, which, again, does not need to be trusted. Here, the detection is on the Qonnector side, which also needs to be able to perform Bell state measurements (BSM) that are inherently probabilistic.
Without quantum memories available at the nodes, the two photons coming from Alice and Bob must arrive at the same time, which lowers the success probability of an MDI-QKD round. The high-level description of the quantum transmission part of the protocol is as follows:

\begin{protocol}{MDI-QKD}
\begin{enumerate}
    \item At each timestep defined by $1/f_{\text{qubit}}$, Alice and Bob both randomly choose a random bit $b\in\{0,1\}$ and a random basis $t \in \{\cross,+\}$ and create qubit $\ket{0}$ if $b=0$ and $t=\cross$, $\ket{1}$ if $b=1$ and $t=\cross$, $\ket{+}$ if $b=0$ and $t=+$ and $\ket{-}$ if $b=1$ and $t=+$.
    \item They both send their state to the Qonnector. 
    \item When the qubits arrives simultaneously, the Qonnector performs a Bell state measurement and communicates the outcome to both parties.
\end{enumerate}

\end{protocol}

This protocol, together with Twin-Field (TF) QKD, which is conceptually close but relies on single-photon instead of two-photon interference, have received tremendous attention in the recent years as they allow for high key rates over record long distances, even beating the repeaterless bound~\cite{PLOB} for the TF case~\cite{Lucamarini_2018,MDIChina,chen2021twinfield}. They are also naturally suited to the progressive vision of quantum networks that can be upgraded as more advanced technology becomes available.
In our analysis, we will provide a basic estimation of the MDI-QKD protocol performance in our Quantum City setting, in particular by considering only the probability of success of the BSM (see Sec.~\ref{sec:model}), but leaving imperfections such as noise or detector dark counts for further analysis. \\

We conclude this section by noting that \textbf{decoy-state BB84}~\cite{BB84coherent1} and Continuous-Variable QKD or \textbf{CV-QKD}~\cite{CVQKD2,CVQKDreview} are also widely studied QKD protocols offering significant advantages for achieving high-performance QKD, including the use of simplified technology, compatible with standard optical communication.
Unfortunately, NetSquid does not support yet models for (weak) coherent state generation or coherent detection techniques used in these protocols, and therefore we do not include them in our analysis. We emphasize, however, that it will be important to develop such network simulation models for a complete analysis of quantum networks, and leave this as an open question.


\subsection{Delegated computation}
\label{sec:Deleg}

One of the most interesting features of the future Quantum Internet is that it will allow users to privately delegate their quantum computation to powerful quantum processors to whom they are connected  \cite{Broadbent_2009, BroadbentDelegated, delegated1}. Recent promising results include protocols providing blind and verifiable delegated computation \cite{lukaSecuring}, even considering classical clients \cite{QFactory} and the corresponding security concerns \cite{LeoSecurity, DelegatedQC}. In such protocols, the quantum processor performing the computation is unaware of what it is computing, and, crucially, the user can also check whether the processor is doing its task correctly.  \\

Delegated quantum computation assumes the existence of a powerful server, that we call here Qomputer, which may work in some computing model, such as, for instance, the measurement-based quantum computation paradigm (MBQC) \cite{MBQC}. Typically, in this case, the server receives quantum states from a client that it uses as input to perform some measurement routine involving back and forth classical communication between server and client. To showcase the feasibility of delegated computation in the context of a Quantum City, we will suppose that a Qomputer operating under the MBQC framework becomes at some point available in our Quantum City. By connecting this Qomputer node to the Qonnector via a quantum channel, any Qlient can securely delegate its computation to this node. We recall here the universal blind delegated protocol from \cite{Broadbent_2009} allowing a Qlient to blindly delegate its computation to a Qomputer node connected to the Qonnector:

\begin{protocol}{Blind delegated computation}
\begin{enumerate}
    \item The Qlient prepares single qubits chosen randomly from $\{1/\sqrt{2}(\ket{0} + e^{i\theta}\ket{1}) \vert \theta = 0, \pi/4, 2\pi/4,...,7\pi/4 \}$ and sends them to the Qonnector.
    \item The Qonnector routes the qubits to the Qomputer.
    \item The Qomputer receives and entangles the qubits in a predefined universal graph state (e.g., the brickwork state from \cite{Broadbent_2009}). 
    \item  Then for each qubit, the Qlient sends a classical message to the Qomputer to tell it in which basis it should measure the qubit. It performs the measurement and communicates the outcome; the Qlient’s choice of angles in future rounds will depend on these values. This interaction continues until all the qubits are measured.
\end{enumerate}

\end{protocol}

The last outcome of the measurements made by the Qomputer then contains the result of the classical function the Qlient is computing. If the Qlient is computing a quantum function, the outcome is the last qubits at the Qomputer node that it can send back to the Qlient. Provided that the Qomputer is able to keep qubits in its memory for a sufficient time, the Qlient only has to manipulate one qubit at a time. In the Quantum City architecture, this makes delegation protocols available to any Qlient. In this work we will not dive into the details of how the qubits are entangled and measured by the Qomputer, but rather estimate the rate at which they can be sent from the Qlient to the Qomputer. This is sufficient to showcase the feasibility of delegated computation protocols when Qomputer nodes become available from a network point of view.


\subsection{Multiparty protocols}
\label{sec:MultiProt}

\begin{figure}[!ht]
    \centering
    \includegraphics[width = 11cm]{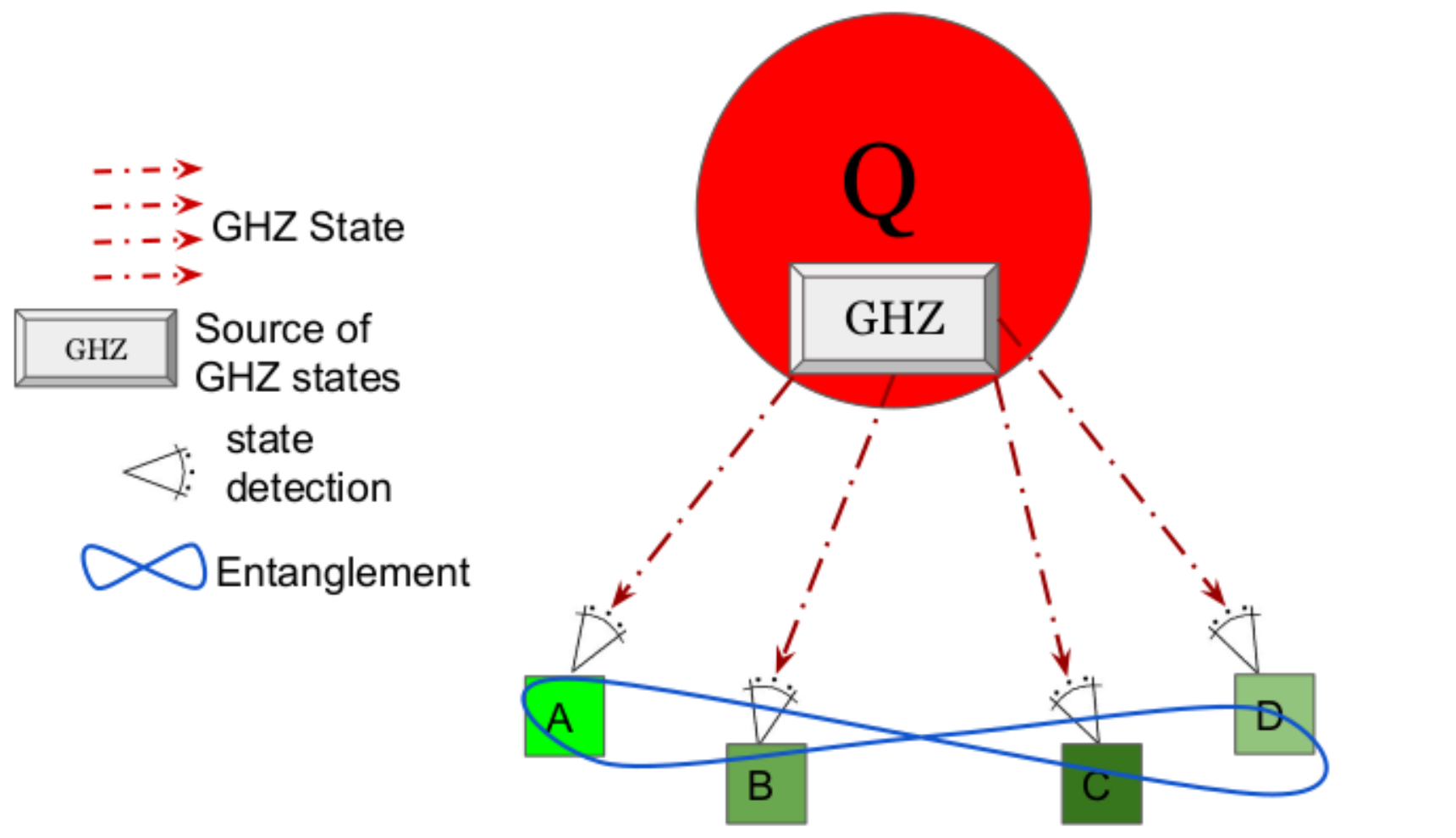}
    \caption{Sharing of a 4-qubit GHZ state from the Qonnector to Qlients.}
    \label{fig:GHZsharing}
\end{figure}

Some of the most groundbreaking applications of quantum networks are multiparty functionalities based on genuine multipartite entanglement shared among $n$ parties~\cite{GME}. In a Quantum City, multiparty protocols that only require the manipulation of a qubit per party at each timestep can be implemented. Here, without loss of generality, we focus on protocols based on the GHZ state, $(\ket{0}^{\bigotimes n}+\ket{1}^{\bigotimes n})/\sqrt{2}$~\cite{GHZ}.
The Qonnector in this case is used as a source of the multipartite GHZ state (see Fig.~\ref{fig:GHZsharing}). Any such protocol thus relies on the ability of the Qonnector node to create these states and to transmit them to the Qlients. As protocols typically rely on sharing many multipartite entangled states sequentially, the rate of creating and sharing these states is a crucial parameter. In this work, we do not look into the details of how GHZ states are created but rather take a high-level view of considering that at a given rate there is a probability $p_{\text{GHZ}-n}$ that an $n$-qubit GHZ state is created (see Sec.~\ref{sec:model}). \\

Two examples of GHZ-based network protocols are conference key agreement \cite{Murta_2020} and anonymous transmission \cite{Anonymity}. The Qlients can choose to trust that their Qonnector is indeed giving them GHZ states or they can incorporate within their protocol entanglement verification rounds~\cite{MEVresistant}. The latter relies on verification protocols that ensure that the source is indeed creating close to GHZ states. We note that these verification protocols consume many states in order to perform the verification task with sufficient confidence. Hence we will see that there is a trade-off between security and rate of performing multiparty protocols. In the following, we describe two multiparty protocols available to Qlients in a Quantum City, as well as a verification protocol with interesting properties.

\subsubsection{Conference Key Agreement}
\label{sec:CKA}
Conference Key Agreement is the multipartite version of QKD. It allows $n$ Qlients to all share a secret key that they can subsequently use for secure communication or other applications. There exists a variety of protocols achieving this functionality (see \cite{Murta_2020} for an extended review) including GHZ-based protocols \cite{Grasselli_2018,CKAPappa}, with a recent experimental realization~\cite{CKAExperimental}. The main ingredient to get a conference secret key in these protocols is GHZ state sharing and measurement. Indeed, an $n$-qubit GHZ state is given by the superposition of all the qubits in the state $\ket{0}$ and all the qubits in the state $\ket{1}$. This means that prior to the measurement of one of the qubits, any qubit measurement outcome is random. However once one qubit has been measured, all the other qubit measurements will give the same output. Hence if a GHZ state is shared to $n$ parties, they all get a private common random bit when measuring their qubit. \\

Similarly to QKD, $n$ Qlients can extract a common secret key from a given raw key shared among them. Here, we will not focus on a specific protocol but simply investigate sharing of $L$ consecutive GHZ states among $n$ Qlients in a Quantum City. This already gives an estimation of the feasibility of conference key agreement protocols in a realistic setting. Just as QKD in the bipartite setting, we use this protocol as a benchmark for multiparty protocols.

\subsubsection{Verification of a GHZ state}
\label{sec:Verif}
We describe below the protocol from \cite{MEVresistant}, which allows $n$ Qlients to verify that the state that they receive is at least $\epsilon$-close to a GHZ state. One Qlient is chosen randomly among them to be the verifier. They have the special role of gathering information on measurement outputs from the other parties. This protocol should be repeated sequentially until sufficient trust is built on the source. The identity of the verifier must be randomized to prevent malicious behavior from the Qlients. In order to prevent a malicious Qonnector to share another state than the GHZ state on the rounds where the Qlients actually use it, the communication or computation rounds have to be randomly chosen among verification rounds.

\begin{protocol}{Multipartite entanglement verification protocol}

\begin{enumerate}
    \item The Qonnector creates an $n$-qubit GHZ state and sends qubit $i$ to party $i$.
    \item The verifier selects for each $i\in[n]$ a random input $x_{i}\in\{0,1\}$ such that $\sum_{i=1}^{n}x_{i}\equiv 0$ (mod 2) and sends it to the corresponding party via an authenticated classical channel. The verifier keeps one to themselves.
    \item If $x_{i}=0$, party $i$ performs a Hadamard operation on their qubit. If $x_{i}=1$, party $i$ performs a $\sqrt{X}$ operation. 
    \item Each party $i$ measures their qubit in the $\{\ket{0},\ket{1}\}$ basis and sends their outcome $y_{i}$ to the verifier via the classical channel.
    \item The verifier accepts and outputs $b_{out}=0$ if and only if 
\begin{equation*}
    \sum_{i=1}^{n}y_{i}\equiv\frac{1}{2}\sum_{i=1}^{n}x_{i} \text{(mod 2)}
\end{equation*}
\end{enumerate}
\end{protocol}

It has been shown that this protocol is secure against any coalition of dishonest Qlients \cite{MEVresistant}, including the Qonnector, and that it is even composably secure in the case where all Qlients are honest, which is often the case when they want to, e.g., get a secure conference key to communicate \cite{Yehia_2021}. Moreover previous works show that by choosing the probability of using the qubit for communication or computation to be $4n\delta/\epsilon^2$, for some $\delta>0 $, all honest parties have the guarantee that the probability the state used has distance more than $\epsilon$ from the correct one is at most $1/\delta$. We have chosen to simulate this verification protocol both because of its good security properties and because it has low hardware requirements for the parties, which matches our assumptions on the Qlients' hardware. Finally, a loss-tolerant version of this protocol has already been implemented with 3 parties~\cite{MEVexperimental}, which shows that it is a good candidate for future quantum network applications.

\subsubsection{Anonymous Transmission}
\label{sec:Anonymous}
This protocol allows two Qlients, a sender and a receiver, to establish a link that they can use for transmitting anonymously a message qubit \cite{Anonymity}. The quantum message is transmitted in a way that the identity of the sender is unknown to every other node, and the identity of the receiver is known only to the sender. It relies on classical pre-processing allowing the sender to notify anonymously the receiver that he is going to receive something. Here we do not describe or simulate this classical pre-processing and we refer the reader to \cite{Anonymity} for more information.

\begin{protocol}{Anonymous Entanglement}
\begin{enumerate}
    \item The Qonnector creates and shares an $n$-qubit GHZ state.
    \item Every Qlient except the sender and the receiver applies a Hadamard gate to their qubit. They measure it and get outcomes $m_i$ that they broadcast.
    \item The sender picks a random bit $b$ and broadcasts it. She applies a $Z$ gate to her qubit if $b=1$.
    \item The receiver picks a random bit $b'$ and broadcasts it. He applies a $Z$ gate to his qubit if $b\oplus\bigoplus_i m_i =1$.
\end{enumerate}
\end{protocol}

After performing this protocol, the sender and the receiver anonymously share a Bell pair that they can use to teleport any qubit state. This relies on the fact that they are able to store their qubit for the duration of the protocol. If on-demand quantum storage is not available at the Qlient nodes, non-optimized solutions like delay lines can be used for this task.  

Repeating the Anonymous entanglement protocol enough times between GHZ verification rounds allows for provably secure anonymous transmission of quantum information. However, the number of required GHZ states does not scale well with the number of parties and the size of the message to transmit. This is because the verification protocol presented above is costly in GHZ resources. We will however see in the next section that it is still possible to perform a few rounds of this protocol in less than an hour with 4 parties and today's hardware capabilities.

\subsection{Figures of merit}
Using a network simulator allows to benchmark different properties of network protocols depending on specific hardware parameters. In this work, we will focus on the \textbf{sifted rate}, in bit per second, at which protocols can be performed. The rate is highly dependent on the quantum state creation rate of the sources, which varies a lot from one setup to another and can usually be tuned to match the single-photon detector's dead time. We will also give an estimation of the \textbf{throughput}, defined as the number of states received divided by the number of states sent (or channel uses). Although they can easily be converted from one to the other given the hardware parameters, the former gives a good estimate of the feasibility of a protocol while the latter characterizes the quality of the qubit transmission in a quantum network. We will also focus on the \textbf{Quantum Bit Error Rate (QBER)} that we define as the number of measurement outputs that where flipped during quantum processes. More explicitly it corresponds to the number of qubits measured in the $\ket{1}$ state where a $\ket{0}$ state was sent (and conversely) over the total number of qubits measured. This bit flipping due to faulty operations during the protocol is a practical measure of the quality of the different pieces of hardware. \\

These parameters allow in most cases to estimate the actual rate at which a protocol can be performed. For example in QKD protocols, the secret key rate is given as a function of the sifted rate and the QBER that depends on the post-processing techniques that is chosen \cite{SecurityQKD}. Here, we focus on the quantum communication and processing parts of protocols, ignoring classical pre and post-processing. Indeed, we simply aim to investigate the performance of a realistic near-term metropolitan-scale quantum network to motivate and guide practical realisations. Based on the results of our simulations, it will be then possible to check in more detail the feasibility of specific protocols.
We hope this will show that the Quantum City is a promising setting, suitable for near-term quantum applications.
\section{Results}
\begin{figure}[!ht]
    \centering
    \includegraphics[width=14.5cm]{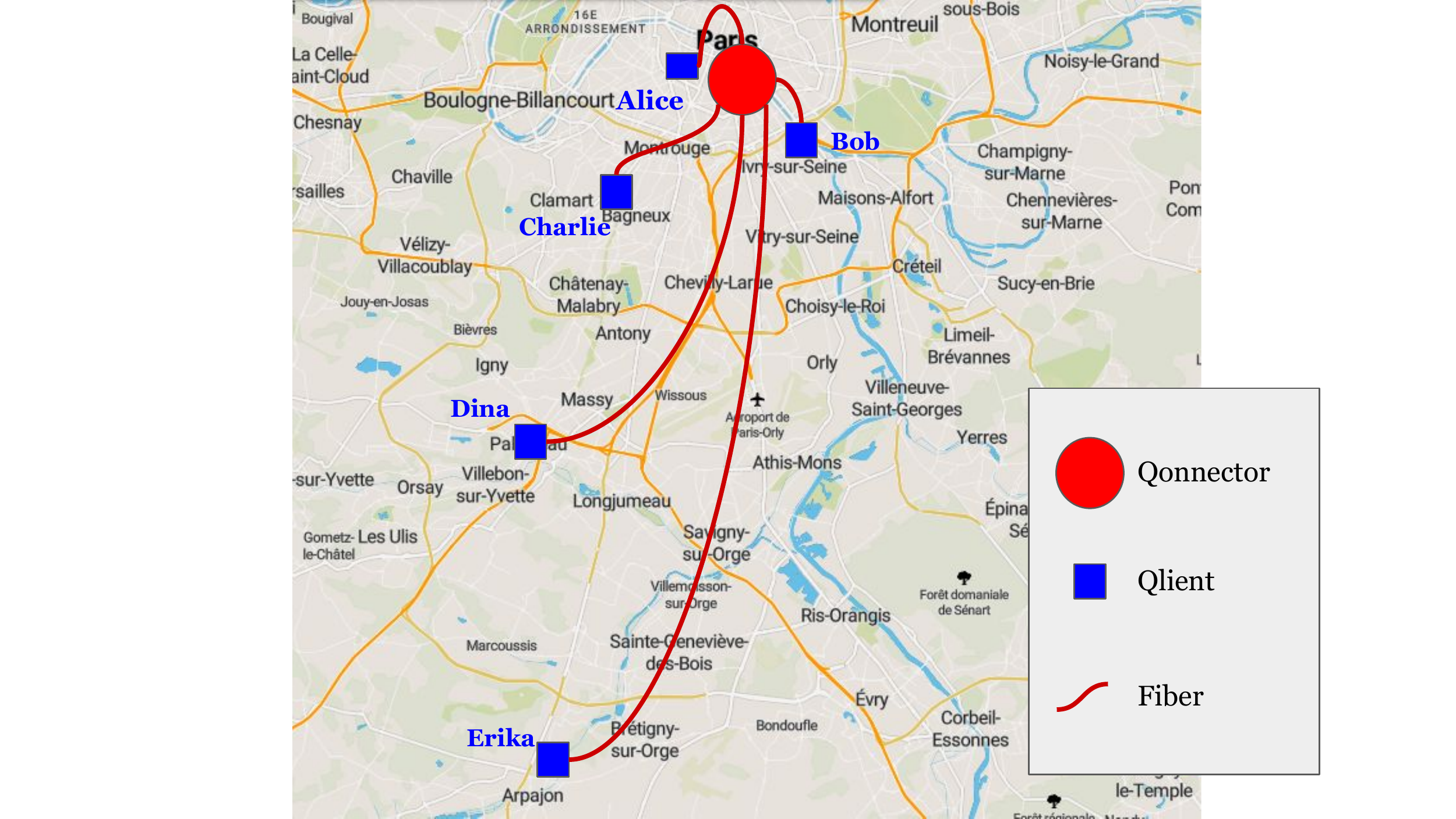}
    \caption{Paris Quantum City: Five Qlients are connected through optical fibers to a Qonnector located at Sorbonne Université (SU) campus. The length of the fiber is 1~m for the link Alice-Qonnector, 3~km for the link Bob-Qonnector, 6~km for the link Charlie-Qonnector, 18~km for the link Dina-Qonnector and 31~km for the link Erika-Qonnector.}
    \label{fig:map}
\end{figure}

We will now simulate a Quantum City in a realistic setting using the simulation tool NetSquid. As explained in Sec.~\ref{sec:model}, errors and losses are modeled with dephasing and depolarising channels that we apply to quantum states when they undergo quantum processes. Our simulation model is formed by five Qlients that represent actual laboratories in the Paris region: Sorbonne Université campus (SU-Alice), Université Paris Cité campus (UPC-Bob), Orange Labs Châtillon (OR-Charlie), Télécom Paris (TP-Dina) and TGCC-CEA (CEA-Erika) (see Fig.~\ref{fig:map}). They are connected through lossy optical fibers to a Qonnector placed in the same lab as Alice. We suppose here that the length of each fiber is given by the line-of-sight distance between the Qonnector and each node. It is easy to see that this choice of placing the Qonnector is not optimal; however, it can allow for more Qlients inside Paris to join and we will see that it already allows for interesting applications.  \\

In NetSquid, protocols are modeled with node subroutines that correspond to the local operations performed by each node during the protocol. By creating the routines that correspond to generating and sending BB84 qubit states, measuring such states in a random basis, transmitting them from one Qlient to another, creating Bell pairs or GHZ states and performing Bell State Measurement, we are able to simulate all the protocols mentioned in Sec.~\ref{sec:QKD}-\ref{sec:MultiProt}, at least from a network point of view. We also created classes of nodes that represent the Qlients and the Qonnector. They contain the necessary classical memory slots for routing and processing outcomes. Note that quantum storage is not included in the routines in this analysis. With these elements we can create a Quantum City instance on which we can simulate the aforementioned protocols.

\subsection{Baseline simulation parameters}
\label{sec:Param}

We start by discussing the set of realistic parameters that we use in our simulations. We consider here photon sources based on spontaneous parametric down conversion (SPDC) in nonlinear crystals, as they are widely used for generation of heralded single-photon and entangled-photon states with high performance in terms of brightness and fidelity~\cite{brunoSource,ChiaraSource}. Such sources are in general cost-effective and can operate at a range of wavelengths. Note however that deterministic single-photon sources, such as those based on semiconductor quantum dots, may also be interesting for quantum network applications~\cite{gao2022,bozzio2022arxiv}.
Our simulations can be adapted to such devices, by suitably changing the parameters. \\

Using SPDC sources, heralded single-qubit states are generated by measuring one photon of a correlated pair with a single-photon detector. EPR pairs are generated with a probability $p_{\text{EPR}}$. 
GHZ states with $2n$ qubits are created by generating $n$ independent EPR pairs and performing on them probabilistic fusion operations~\cite{EntanglementFusionXP}. We assume here that the probability of success of the fusion operation is the same as the one of a linear-optic BSM~\cite{BSMmaxProba} with some additional losses, and set it to $p_{\text{fusion}} = 0.36$.
We can then deduce the average generation rate for $2n$-qubit GHZ states:
\begin{equation}\label{eq:RateEven}
    GR_{\text{GHZ}-2n} = f \cdot p_{\text{EPR}}^n \cdot p_{\text{fusion}}^{n-1},
\end{equation}
where $f$ is the pump laser repetition rate and we consider each laser pulse as an attempt to generate a state. \\
This way, we can evaluate the probability of generating a GHZ state as:
\begin{equation}\label{probaGHZGenEven}
p_{\text{GHZ}-2n} = p_{\text{EPR}}^n \cdot p_{\text{fusion}}^{n-1}.   
\end{equation}
GHZ states with $(2n-1)$ qubits are generated by measuring a heralding qubit of a $2n$-qubit GHZ state, resulting in an average generation rate:
\begin{equation}\label{eq:RateEven}
    GR_{\text{GHZ}-(2n-1)} = GR_{\text{GHZ}-2n} \cdot \eta_{\text{herald}} = f \cdot p_{\text{EPR}}^n \cdot p_{\text{fusion}}^{n-1} \cdot \eta_{\text{herald}},
\end{equation}
where $\eta_{\text{herald}}$ is the probability of measuring the heralding photon, including detector and coupling efficiencies, as well as any losses in optical components. For an optimized setup with high-efficiency detectors, we can consider $\eta_{\text{herald}} \simeq 0.7-0.8$. Similarly to the $2n$-qubit GHZ state case, we can then evaluate the probability of generating a single photon or a GHZ state in a laser pulse as: 
\begin{equation}\label{probaGHZGenOdd}
\begin{aligned}
p_{\text{qubit}} &= p_{\text{EPR}} \cdot \eta_{\text{herald}}\\
p_{\text{GHZ}-(2n-1)} &= p_{\text{EPR}}^n \cdot p_{\text{fusion}}^{n-1}\cdot \eta_{\text{herald}}
\end{aligned}
\end{equation}

Most current experiments use lasers that do not exceed a pulse repetition rate of $f = \SI{80}{\mega\hertz}$. Temporal multiplexing can be used to increase the average rate of emission while keeping the noise low~\cite{ChiaraSource}; however, we wish to keep $f\cdot p_{\text{EPR}} \leq \SI{10}{\mega\hertz}$, as a higher pair emission rate would lead to a drop of the detector performance because of the recovery time, which is typically $\lesssim \SI{100}{\nano\second}$. Hence, we take $f = \SI{80}{\mega\hertz}$ and $p_{\text{EPR}}=0.01$ for $1$ and $2$ qubit experiments, in order to limit the noise due to double emission \cite{CombJin}. For experiments with more photons, we take a higher value $p_{\text{EPR}} = 0.1$ in order to favor multiple pair emission in one pulse, while keeping a lower $f=\SI{8}{\mega\hertz}$.\\

Below we list the parameters used for the simulations. Most of them are parameters witnessed in experiments today, or are derived from what is expected to be possible in the near term. Some others, such as the error probabilities $p_{\text{flip}}$, $p_{\text{crosstalk}}$ or $p_{\text{dephase}}$, highly depend on how optimized the experiment is. We therefore choose somewhat arbitrary parameters that can be easily modified in our code in order to simulate errors in the protocols. 
We also set somewhat arbitrarily the routing probability $p_{\text{transmit}}$, leaving the possibility to change it in order to model novel techniques. Finally, we set the parameters $p_{\text{det}}$, $R_{\text{dark}}$, and $\Delta t_{\text{det}}$ to values corresponding to high-performance superconducting nanowire single-photon detectors, such as those commercially available, for instance, by ID Quantique.\\

\textbf{Parameters:}\\ 

\begin{tabular}{l|c|l}
    $f_{\text{qubit}}$ & $\SI{80}{\mega\hertz}$ & Qubit creation attempt frequency\\
    $p_{\text{qubit}}$ & ${8\cdot10^{-3}}$  & Success probability of creation of a qubit  \\
    $p_{\text{flip}}$ & 0 & Flipping probability at the creation of a qubit \\
    $p_{\text{crosstalk}}$ & $10^{-5}$ & Probability that the detector flips the outcome \\
    $f_{\text{EPR}}$ & $\SI{80}{\mega\hertz}$ & EPR pair creation attempt frequency \\
    $p_{\text{EPR}}$ & $10^{-2}$ & Success probability of the creation of an EPR pair \\
    $p_{\text{BSM}}$ & 0.36 & probability that a Bell state measurement succeeds \\
    $f_{\text{GHZ}}$ & $\SI{8}{\mega\hertz}$ &  GHZ state creation attempt frequency\\
    $p_{\text{GHZ}-3}$ & $2.5\cdot 10^{-3}$ & Probability that an attempt of a 3-qubit GHZ state creation succeeds \\
    $p_{\text{GHZ}-4}$ & $3.6\cdot 10^{-3}$ & Probability that an attempt of a 4-qubit GHZ state creation succeeds \\
    $p_{\text{GHZ}-5}$ & $9\cdot 10^{-5}$ & Probability that an attempt of a 5-qubit GHZ state creation succeeds \\
    $p_{\text{transmit}}$ & 0.9 & Probability that routing a qubit succeeds \\
    $t_{\text{gate}}$ & $\SI{1}{\nano\second}$& Time it takes to perform an operation on one qubit \\
    $p_{\text{coupling}}$ & 0.9 & Fiber coupling efficiency \\
    $\eta_{\text{fiber}}$ & $\SI{0.18}{\dB/\kilo\meter}$ & Fiber loss per kilometer \\ 
    $p_{\text{dephase}}$ & 0.02 & Phase flip probability in the fiber  \\ 
    $p_{\text{det}}$ &  0.95 & Detector efficiency (probability that a measurement succeeds)  \\
    $R_{\text{dark}}$ & $10^2\si{\hertz}$ & Dark count rate \\
    $\Delta t_{\text{det}}$ & $\SI{100}{\pico\second}$ & Detector detection gate \\
\end{tabular}\\

\newpage
\subsection{Bipartite protocols }

\subsubsection{Simulation of QKD protocols}

We start our network simulation analysis with the performance of the Quantum City architecture in the setting of Fig.~\ref{fig:map} for Quantum Key Distribution. As discussed in Sec.~\ref{sec:QKD}, there are many protocols achieving this functionality, which differ in the involved hardware and in the trust that Qlients give to the Qonnector. Here we suppose that each Qlient node is capable of manipulating (creating and/or measuring) one qubit at a time. The Qlients choose among the different protocols depending on their hardware or on which feature is more desired.\\ 

We perform between 200 and 500 simulation runs for each protocol. 
By averaging over these runs, we give an estimate of the sifted key rate by dividing the length of the sifted key over the simulation time. We also estimate the throughput by dividing the number of photons received by the number of photons sent as well as the QBER by counting the number of bits that have been flipped at the end of protocol. Finally, we show plots with the accumulated sifted key after a certain simulation time. The sifted key rate and QBER are the two main parameters that are used for the calculation of the secret key rate of QKD protocols.\\

\textbf{BB84.} We simulate the performance of the BB84 protocol in the two settings described earlier: between the Qonnector and each of the Qlients (corresponding to the case 2xBB84 and 2xBB84 reversed in Fig.~\ref{fig:AllQKD}) and between two Qlients using the Qonnector as a router (corresponding to the case transmitted BB84 in Fig.~\ref{fig:AllQKD}). 
In the first case, we assume for the moment that all the nodes have the same hardware parameters so it does not matter whether the sender is the Qonnector or the Qlient. In the second case, the photon transmission at the Qonnector node succeeds with probability $p_{\text{transmit}}=0.9$; we recall that this routing parameter can easily change in the simulation.\\ 

The simulation is performed as follows: At each timestep defined by $1/f_{\text{qubit}}$ as the time necessary to create a BB84 state, a qubit is created at a sending node. The associated classical bit as well as the basis and a timestamp are stored in classical memory slots. Qubit states are sent through the fiber channels along with a classical message containing the timestamp. They are then measured at the receiving node and outcomes are stored in classical memory slots alongside the measurement basis and the timestamp.  After a fixed simulation time, we perform sifting on the two resulting lists, using the timestamps to compare the measurement bases. This leaves us with correlated lists of raw key bits at the sending and receiving nodes, from which we can extract the data of interest for our analysis.\\

\begin{figure}[!ht]
    \centering
    \includegraphics[width=14cm]{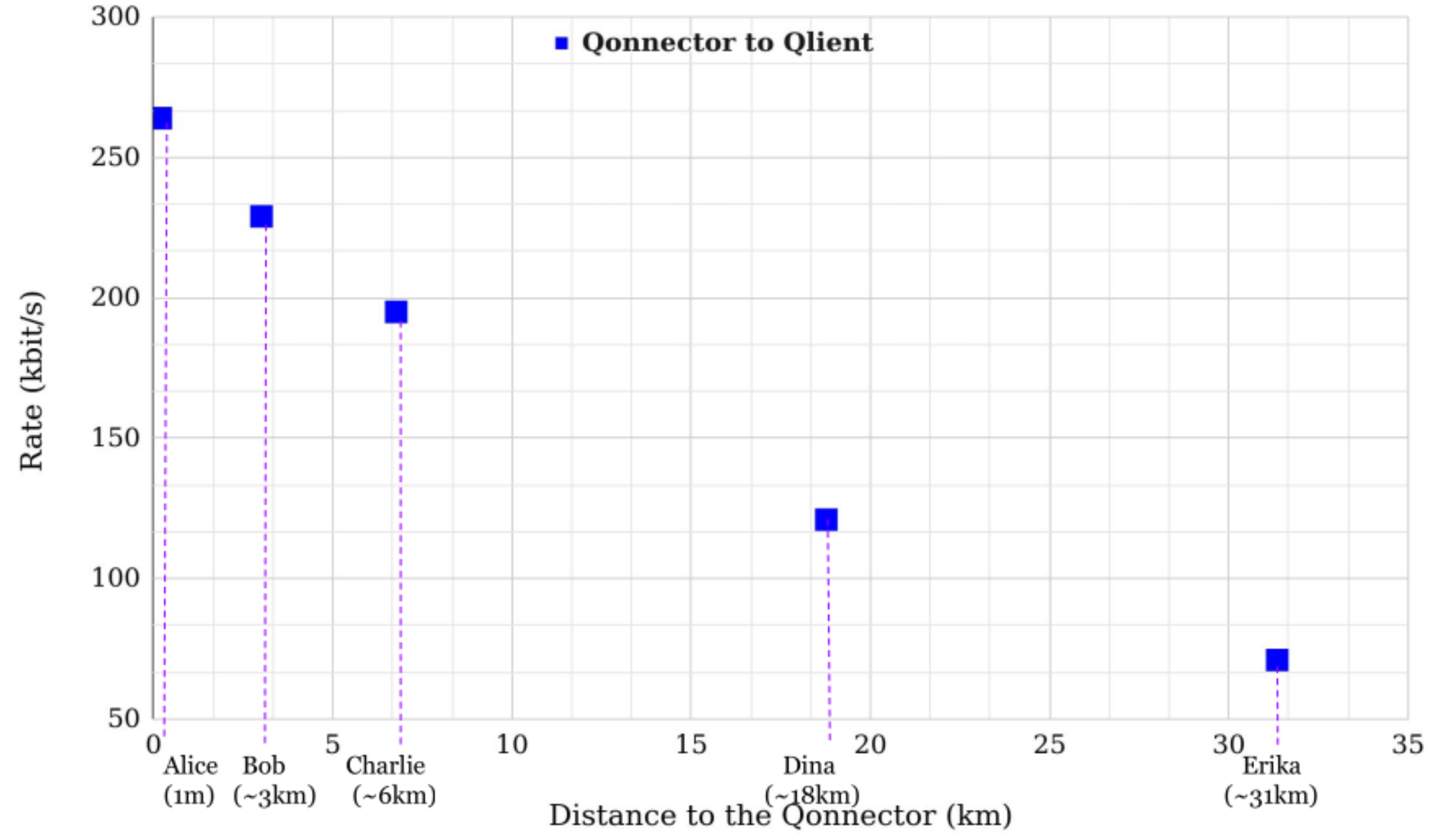}
    \caption{Rate of BB84 state transmission from the Qonnector to the Qlients (blue squares). The rate here represents the number of sifted key bits per second shared between each Qlient and the Qonnector.}
    \label{fig:BB84SimpleRate}
\end{figure}

\begin{figure}[!ht]
    \centering
    \includegraphics[width=12cm]{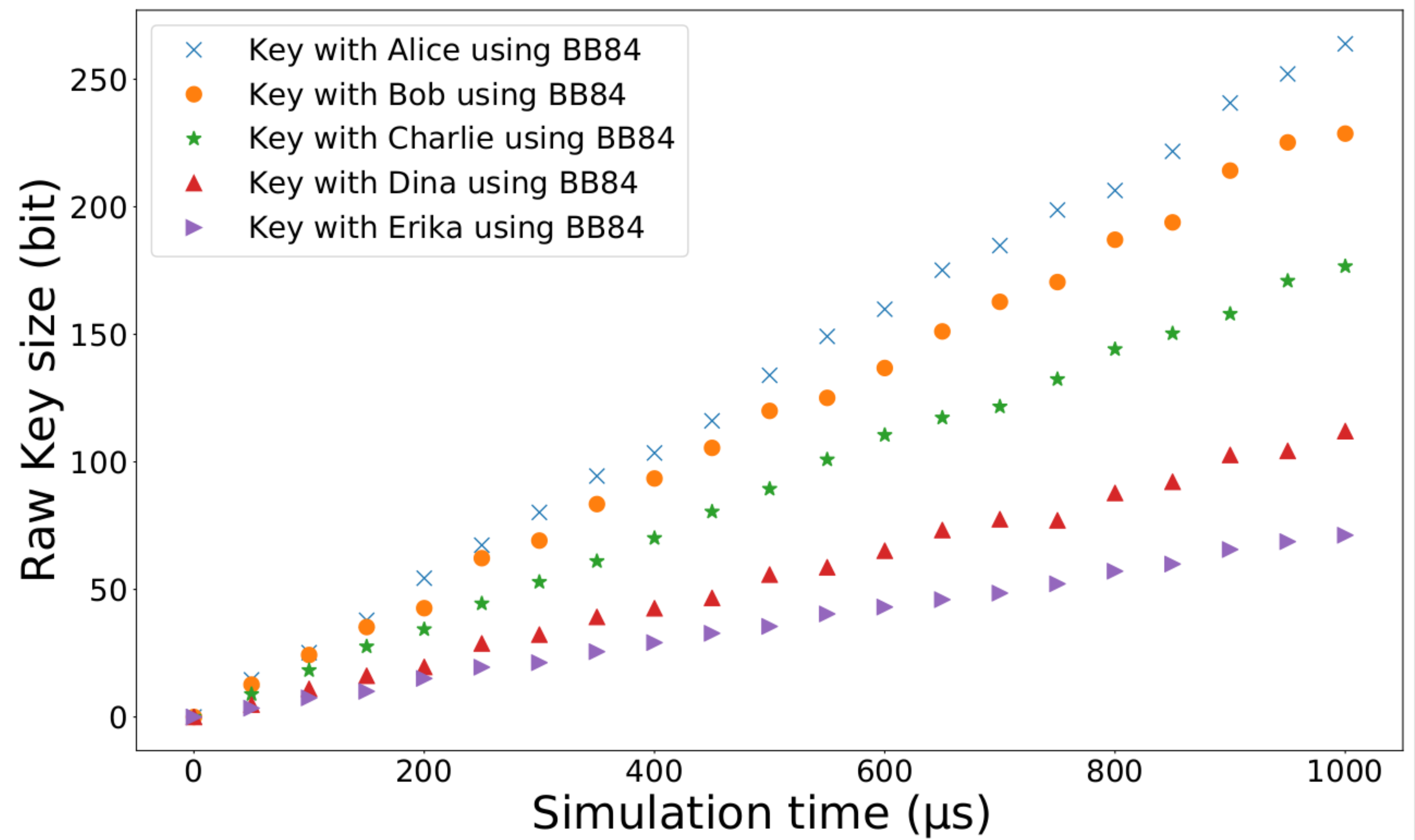}
    \caption{Comparison between the number of successful BB84 qubit transmissions from the Qonnector to every Qlient of our network. }
    \label{fig:BB84comparison}
\end{figure}

\begin{figure}[!ht]
    \centering
    \includegraphics[width=12cm]{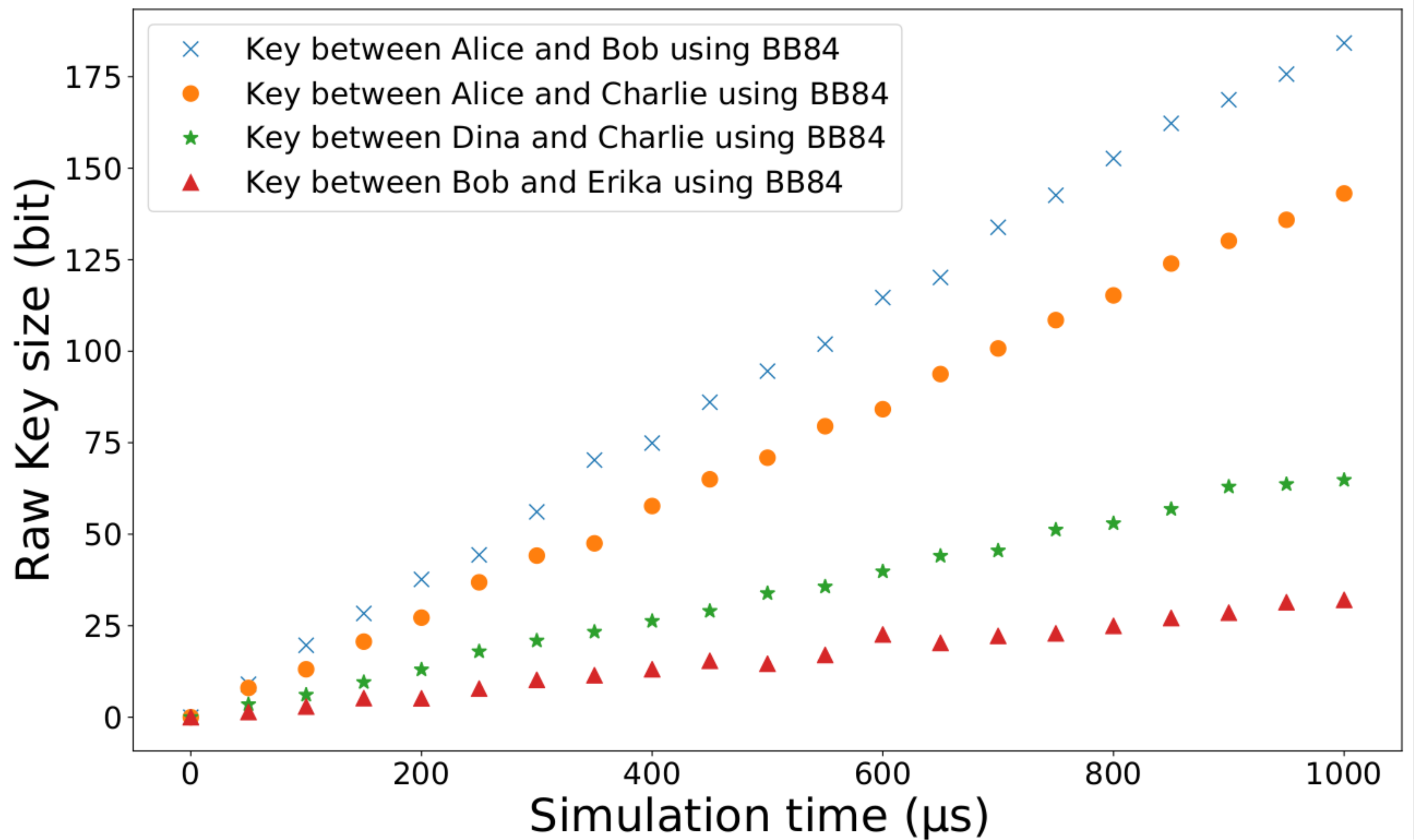}
    \caption{Comparison between the number of successful BB84 qubit transmissions from some Qlients to others through the Qonnector.}
    \label{fig:BB84trans}
\end{figure}

In Table~\ref{tab:BB84forth} we show the achieved sifted key rate, throughput and QBER after a few hundred rounds of simulation. We also show in Fig.~\ref{fig:BB84SimpleRate} the rate for each Qlient as a function of the distance from the Qonnector and in Figs.~\ref{fig:BB84comparison} and \ref{fig:BB84trans} the number of successful BB84 rounds as a function of the simulation time for each setting.\\ 

\begin{table}[!ht]
    \centering
    \begin{tabular}{|c|m{3cm}|m{3.5cm}|c|}
    \hline
         Nodes involved & Rate (sifted key bit per second) & Throughput (sifted key bit per channel use) & QBER \\ \hline
         Qonn $->$ Alice & 263900 & 0.423 & 1.0\% \\
         Qonn $->$ Bob & 228700 & 0.374 & 0.9\% \\
         Qonn $->$ Charlie & 200700 & 0.322 & 1.0\% \\
         Qonn $->$ Dina & 116850 & 0.180 & 0.9\% \\
         Qonn $->$ Erika & 71250 & 0.115 & 0.9\% \\ \hline
         Alice $->$ Qonn $->$ Bob  & 184200 & 0.2185 & 1.8\% \\
         Alice $->$ Qonn $->$ Charlie & 158450 & 0.2592 & 1.8\% \\
         Dina $->$ Qonn $->$ Charlie & 72700 & 0.1078 & 1.9\% \\
         Bob $->$ Qonn $->$ Erika & 51950 & 0.0845 & 1.7\% \\\hline
    \end{tabular}
    \caption{Performance of the BB84 protocol in the Paris Quantum City. The first five lines correspond to the Qonnector sending BB84 states to each Qlient, and the last four correspond to pairs of Qlients using the Qonnector as a transmitting station.}
    \label{tab:BB84forth}
\end{table}

The actual rate that corresponds to sharing a key between two Qlients using the BB84 protocol between the Qonnector and the Qlients is given by the minimum of the key rates with each individual Qlient. 
As expected, photon loss in the fiber affects directly the performance of this protocol. We can see the rate dropping for nodes situated further away from the Qonnector, dropping even more when the Qonnector routes a photon coming from one Qlient to another. Despite this lower performance, Qlients do not have to trust the Qonnector in the latter case.\\

\newpage
\textbf{Entanglement-based QKD.} We now investigate the performance of the Quantum City for QKD when EPR pairs are sent between Qlients. We simulate the following BBM92 scenario: at each timestep defined by $1/f_{\text{EPR}}$ an EPR pair is created and sent by the Qonnector to two Qlients who measure it in randomly selected basis according to the protocol. As in the BB84 simulations, outcomes, timestamps and measurement bases are stored in classical memory slots. We can then use the resulting lists to extract relevant data. Note that to estimate the QBER we count the timesteps where Qlients obtained correlated results.\\

\begin{figure}[!ht]
    \centering
    \includegraphics[width=12cm]{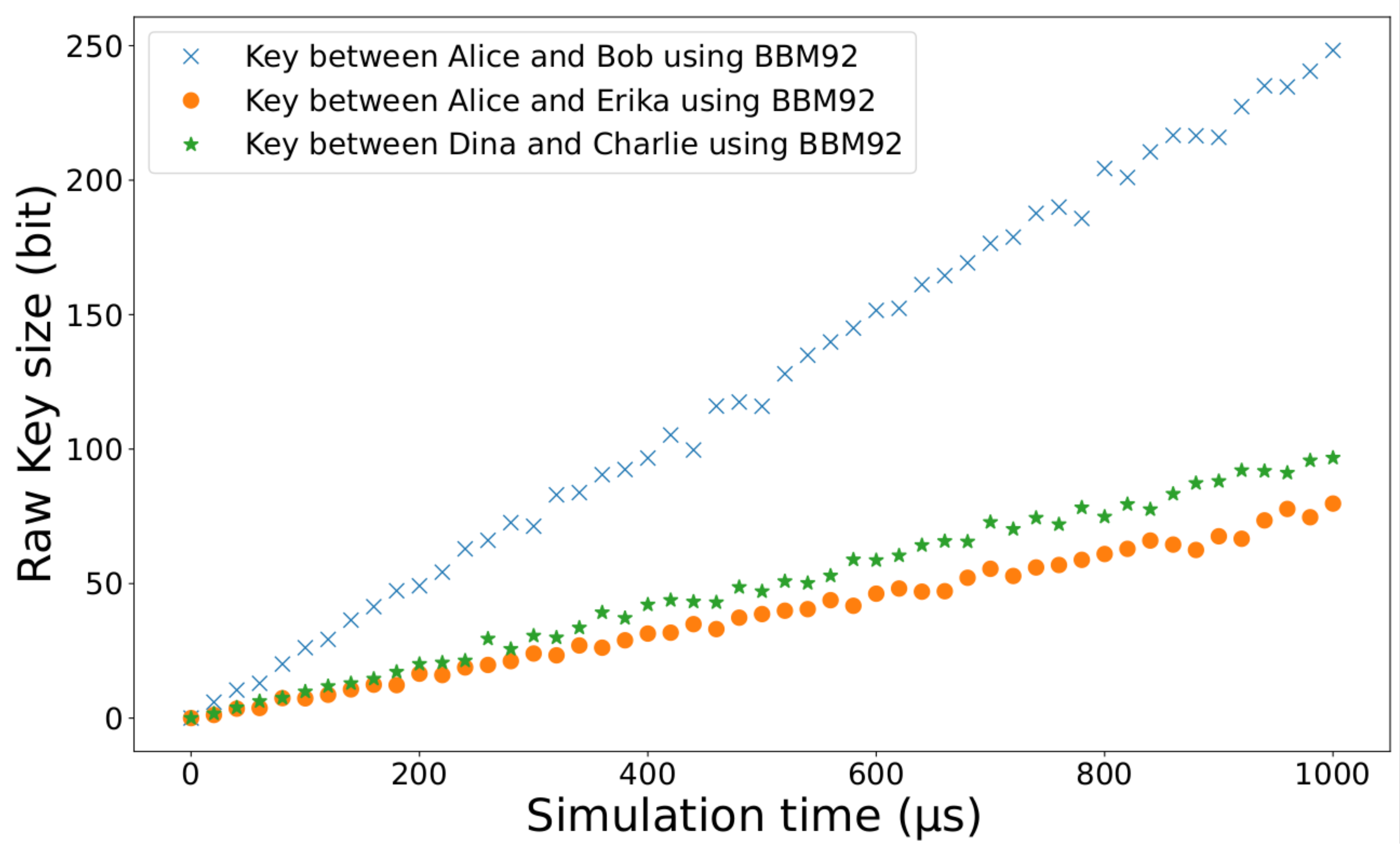}
    \caption{Comparison between the number of EPR pairs successfully measured by pairs of Qlients.}
    \label{fig:EPR}
\end{figure}

In Table~\ref{tab:EPR} we show the EPR state sharing rate and throughput as well as the QBER for a few pairs of Qlients. We also plot in Fig.~\ref{fig:EPR} the accumulated sifted key obtained by pairs of Qlients receiving and measuring EPR pairs from the Qonnector and counting the qubits received with the same timestamp and measured in the same basis. These results are consistent with real field results, for example those obtained in the QKD testbed deployed in the city of Nice~\cite{NiceNetwork}.\\

\begin{table}[!ht]
    \centering
    \begin{tabular}{|c|m{3cm}|m{3.5cm}|c|}
    \hline
         Nodes involved & Rate (EPR pairs per second) & Throughput (pairs received over pairs sent) & QBER \\ \hline
         Alice $<-$ Qonn $->$ Bob  & 248250 & 0.2068 & 1.9\% \\
         Alice $<-$ Qonn $->$ Erika & 79750 & 0.1042 & 1.3\% \\
         Dina $<-$ Qonn $->$ Charlie & 96750 & 0.1252 & 2.2\% \\\hline
    \end{tabular}
    \caption{Performance of the BBM92 protocol between pairs of Qlients.}
    \label{tab:EPR}
\end{table}

Note that the EPR sharing rate also showcases how long it takes to create entanglement between two nodes in the Quantum City. As explained in the introduction, this is a crucial characteristic of a quantum network. To consider this properly, it is necessary to upgrade the Qlient capacities with quantum storage. In addition to allowing for efficient, on-demand operations, this also opens the way to a whole range of applications based on teleportation and entanglement swapping techniques.\\

\textbf{MDI-QKD.} Finally we simulate the MDI-QKD protocol as follows: at each timestep the Qlients prepare and send BB84 states to the Qonnector who performs a Bell state measurement on them. If the measurement is successful, the outcome along with its timestamp is stored in a classical memory slot. We do not simulate classical post-processing and consider that a round is successful when the joint BSM is successful. We show the performance of MDI-QKD between two different Qlients in Table~\ref{tab:MDI}. \\ 

\begin{table}[!ht]
    \centering
    \begin{tabular}{|c|m{3.8cm}|m{3.8cm}|}
    \hline
         Qlient involved & Rate (successful MDI QKD rounds per second) & Throughput (successful rounds over photons sent) \\ \hline
         Alice $->$ Qonn $<-$ Bob  & 420 & 0.05\% \\
         Alice $->$ Qonn $<-$ Charlie & 330 & 0.04\% \\
         Dina $->$ Qonn $<-$ Charlie & 240 &  0.03\% \\ 
         Bob $->$ Qonn $<-$ Erika & 30 & 0.004\%\\\hline
    \end{tabular}
    \caption{Performance of the MDI-QKD protocol between pairs of Qlients.}
    \label{tab:MDI}
\end{table}

The low success probability of the BSM combined with the low probability that both photons from the Qlients arrive at the same time at the Qonnector explains the lower rate of this scheme compared to the previously studied schemes.
The performance can be greatly improved by using quantum memories, which would store the qubit arriving from one Qlient until a qubit arrives from the other Qlient~\cite{repeaterNV}.
We also recall that in MDI-QKD the Qlients do not need to trust the Qonnector and the single-photon detectors are on the Qonnector side, which is favorable in terms of required resources at the Qlient nodes.\\


\subsubsection{Optimizing the parameters}

To test our network architecture in the most realistic setting, we can choose different capabilities for each node. As an example let us set Bob to have the best detector parameters but a low-performance transmitter ($p_{\text{qubit}}=5\cdot10^{-3} $ and $p_{\text{flip}} = 0.01 $) and Dina to have the best transmitter parameters but low-performance detectors ($p_{\text{det}}=0.85$ , $p_{\text{crosstalk}}=10^{-2}$, $R_{\text{dark}}=10^4\si{\hertz}$ and $\Delta t_{\text{det}}=\SI{500}{\pico\second}$). Moreover, let Charlie represent the most limited Qlient with the lowest abilities both in sending and detecting states. We leave the Qonnector as well as Alice and Erika to have the best possible choice of parameters. We emphasize that all these parameters can be easily modified in our code and that the simulation modules are available on GitHub~\cite{github}. In this section we will refer to this more realistic set of parameters as the modified set of parameters. \\

\begin{figure}[!ht]
    \centering
    \includegraphics[width=12cm]{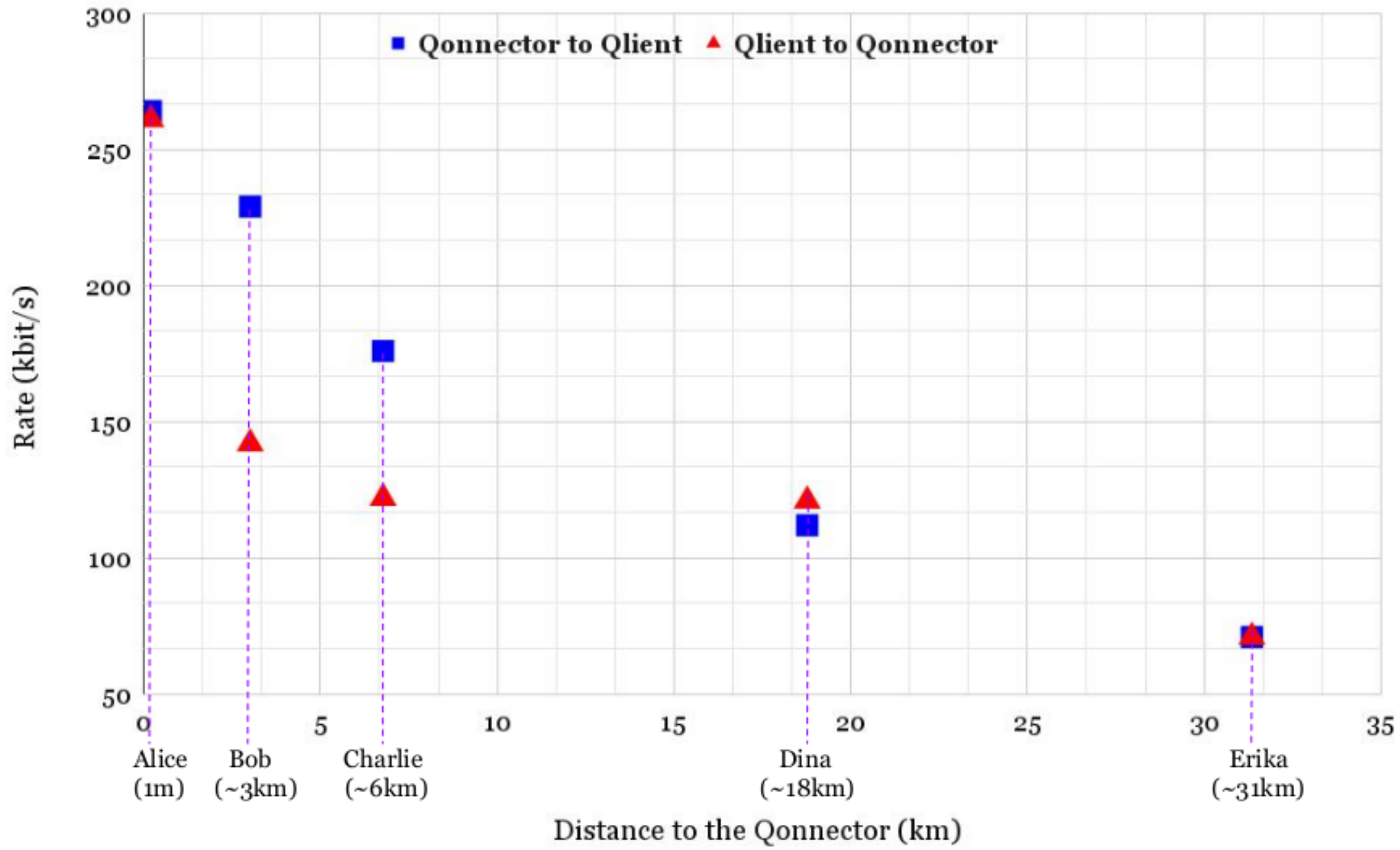}
    \caption{Rate (in sifted key bits per second) of BB84 state transmission from the Qonnector to the Qlients (blue squares) and from the Qlients to the Qonnector (red triangles) with the modified set of parameters. }
    \label{fig:BB84rate}
\end{figure}
\vspace{-0.2cm}

In Table~\ref{tab:BB84backandforth} we show the sifted key rate, throughput and QBER for sending and receiving BB84 states both ways between the Qonnector and each Qlient. We also plot the rate as a function of the distance between the Qonnector and each Qlient (see Fig.~\ref{fig:BB84rate}). We see that the different quality in hardware directly reflects on the simulations outcomes. For example Bob who has good detection but poor transmission capabilities, performs better as a receiving node and Dina, with the reverse capabilities, performs better as a sending node. We remark that with the simulation parameters we have chosen, the effect of the transmitting capability is more pronounced that the detection one.\\

\begin{table}[!ht]
    \centering
    \begin{tabular}{|c|m{3cm}|m{3.5cm}|c|}
    \hline
         Nodes involved & Rate (sifted key bit per second) & Throughput (sifted key bit per channel use) & QBER \\ \hline
         Qonn $->$ Alice & 263900 & 0.4233 & 1.0\% \\
         Qonn $->$ Bob & 228700 & 0.3742 & 0.9\% \\
         Qonn $->$ Charlie & 175650 & 0.2864 & 1.9\% \\
         Qonn $->$ Dina & 112050 & 0.1804 & 2.2\% \\
         Qonn $->$ Erika & 71250 & 0.1156 & 0.9\% \\ \hline
         Alice $->$ Qonn & 260450 & 0.4323 & 1.0\% \\
         Bob $->$ Qonn & 141800 & 0.3735 & 2.0\% \\
         Charlie $->$ Qonn & 122400 & 0.3218 & 2.0\% \\
         Dina $->$ Qonn & 121750 & 0.1963 & 1.0\% \\
         Erika $->$ Qonn & 70750 & 0.1142 & 0.9\% \\ \hline
    \end{tabular}
    \caption{Performance of the BB84 protocol between Qlients and the Qonnector in the Paris Quantum City with the modified set of parameters.}
    \label{tab:BB84backandforth}
\end{table}

\begin{figure}[!ht]
    \centering
    \includegraphics[width=10cm]{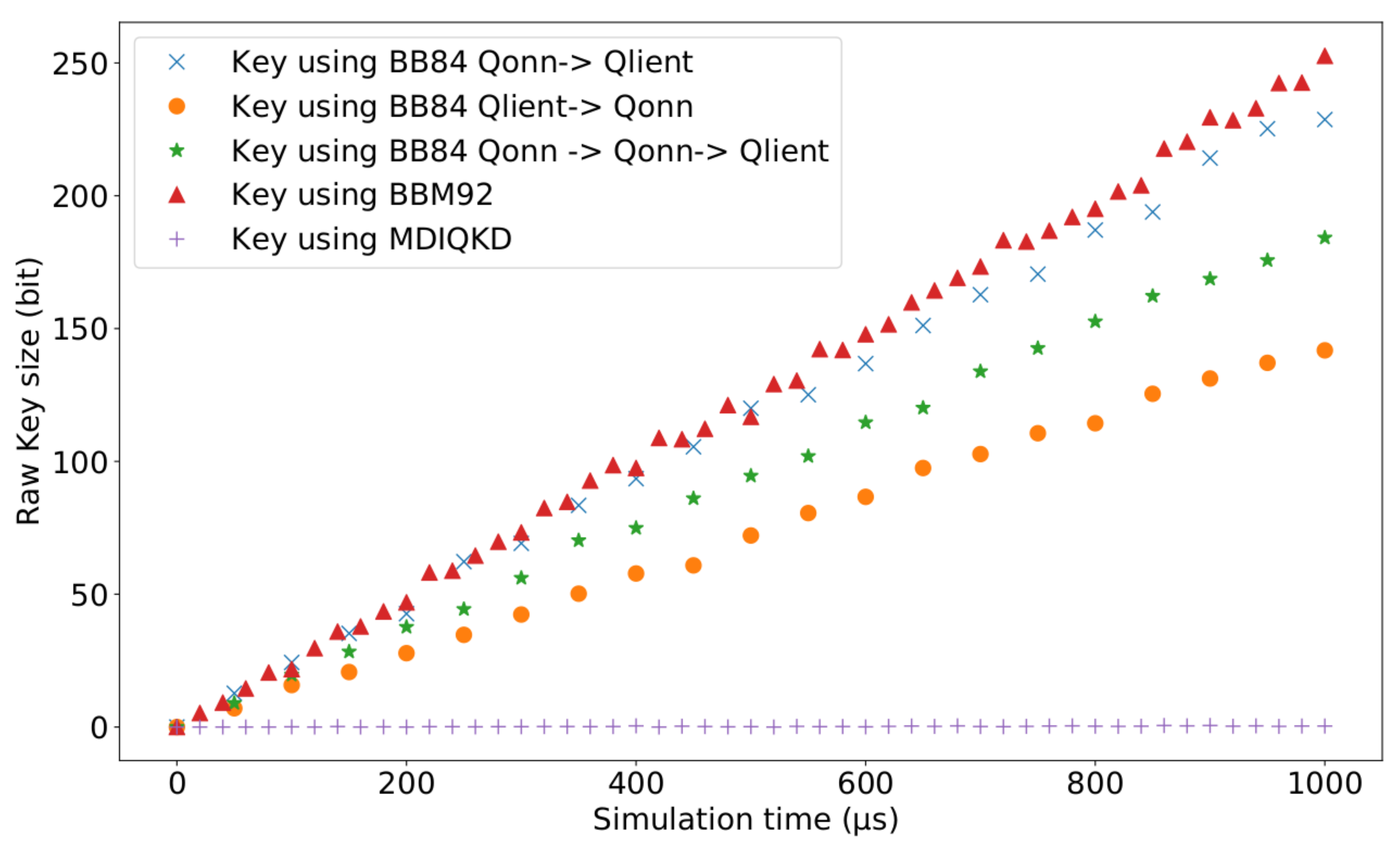}
    \caption{Comparison between different QKD protocols for Alice and Bob. We show the size of the sifted key shared between the two Qlients as a function of the simulation time.}
    \label{fig:ABQKDComp}
\end{figure}

\begin{figure}[!ht]
    \centering
    \includegraphics[width=10cm]{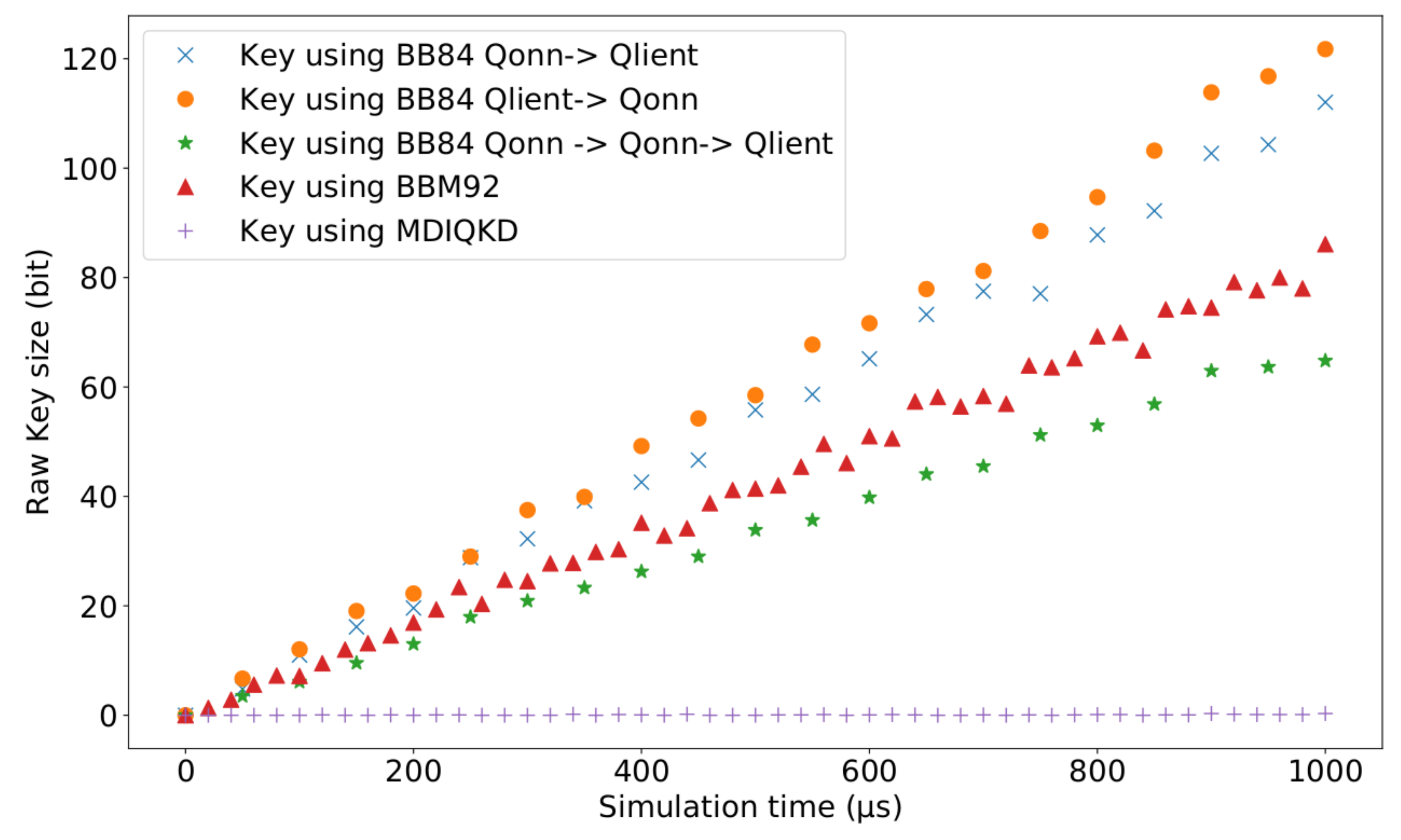}
    \caption{Comparison between different QKD protocols for Charlie and Dina. We show the size of the sifted key shared between the two Qlients as a function of the simulation time. }
    \label{fig:CDQKDCompare}
\end{figure}

We have studied the performance by simulating different QKD protocols in a realistic Quantum City network setting. With simple optical elements on the Qlient side, we have seen various ways for two Qlients to secretly share a key that we can now compare. In Figs.~\ref{fig:ABQKDComp} and \ref{fig:CDQKDCompare} we show the accumulated sifted key length as a function of the simulation time for different QKD protocols for Alice and Bob and for Charlie and Dina, with the modified set of parameters. With these parameters, we can see that Alice and Bob can use favorably entanglement-based QKD while BB84 from the Qlients to the Qonnector is optimal for Charlie and Dina. For these small simulation times, MDI-QKD generates very few bits of shared key due to its very low success probability, which appears as zero when averaging over all the simulation runs. This highlights that such network simulation tools can allow in a fast and flexible way for resource optimization when choosing between different protocols for a target functionality, while also considering the trade-off between performance and desired trust requirements.

\newpage
\subsubsection{Delegated computation}

Let us now imagine that one of the nodes of our quantum City, say Erika, grows to have the abilities of a quantum computer and becomes a Qomputer node. As explained in Sec.~\ref{sec:Deleg}, by simply sending single qubits and then classically communicating with Erika, any Qlient of the Quantum City can enjoy Erika's quantum computing power. In Table~\ref{tab:Deleg} we show the rate at which single qubits can be sent from each of the Qlients to Erika with our baseline set of parameters.\\

\begin{table}[!ht]
    \centering
    \begin{tabular}{|c|c|}
    \hline
         Qlient  & Rate (successful photon transmissions per second)  \\ \hline
         Alice   & 118200  \\
         Bob & 64940  \\
         Charlie & 54480  \\ 
         Dina & 53520 \\\hline
    \end{tabular}
    \caption{Performance of qubit transmission from each Qlient to Erika through the Qonnector.}
    \label{tab:Deleg}
\end{table}

We do not know a priori how this rate exactly relates to the rate of the actual computation a Qlient could perform, due to the unknown capabilities of the future Qomputer node. However, the work from \cite{Mantri_2013} shows that the method that we have described in Sec.~\ref{sec:Deleg} comes within a factor of $8/3$ of being optimal in the used resources. This means that in principle, new developments could reduce the number of required qubits to remotely perform an operation. The optimization of delegation protocols, which also depend on the Qomputer technology and computing model, is outside the scope of this work. Note also that according to the results from \cite{QFactory}, a Qlient could even securely use the Qonnector to send qubits to Erika and delegate a quantum computation while being completely classical.


\subsection{Multiparty protocols}

As discussed in Sec.~\ref{sec:MultiProt}, multiparty protocols rely on multipartite entangled states shared between the users of the network. Restricting our network simulation analysis to GHZ states, the relevant figure of merit for assessing the performance in this case is the rate of successful transmission of such states. Protocols using other states are not very different from a network simulation point of view; the main difference would be the probability that they are successfully created. \\

Following the analysis of Sec.~\ref{sec:Param}, we assume that $n$-qubit GHZ states are generated at the Qonnector node at a rate $f_{\text{GHZ}}$ with a probability $p_{\text{GHZ}-n}$. The qubits are then sent through the channels to the Qlients, who record the number of detection events. Precise synchronization is required for correctly assessing the obtained correlations. Here for simplicity, we just consider the events that correspond to the same timestamp.\\

We present in Table~\ref{tab:GHZ} the estimated GHZ sharing and error rates in our Quantum City setting with the baseline set of parameters. The error rates have been estimated by counting the number of GHZ states successfully shared in which at least one of the outcome bits has been flipped during the process. For GHZ-5 the number of successful GHZ state transmissions is too low to have a correct estimation of the error rate.\\

\begin{table}[!ht]
    \centering
    \begin{tabular}{|c|m{3.8cm}|m{3.8cm}|}
    \hline
         Qlients involved & Rate (successful GHZ states shared per second) & GHZ error rate \\ \hline
         GHZ3 to Alice, Bob and Charlie  & 4260 & 2.1\% \\
         GHZ4 to Alice, Bob, Charlie and Dina & 4495 & 1.8\% \\
         GHZ5 to Alice, Bob, Charlie, Dina and Erika & 45 &  - \\ \hline
    \end{tabular}
    \caption{Performance of GHZ sharing from the Qonnector to 3, 4 and 5 Qlients. These results have been obtained after averaging over 500 runs of $2000$~$\mu s$.}
    \label{tab:GHZ}
\end{table}

\begin{figure}[!ht]
    \centering
    \includegraphics[width=12cm]{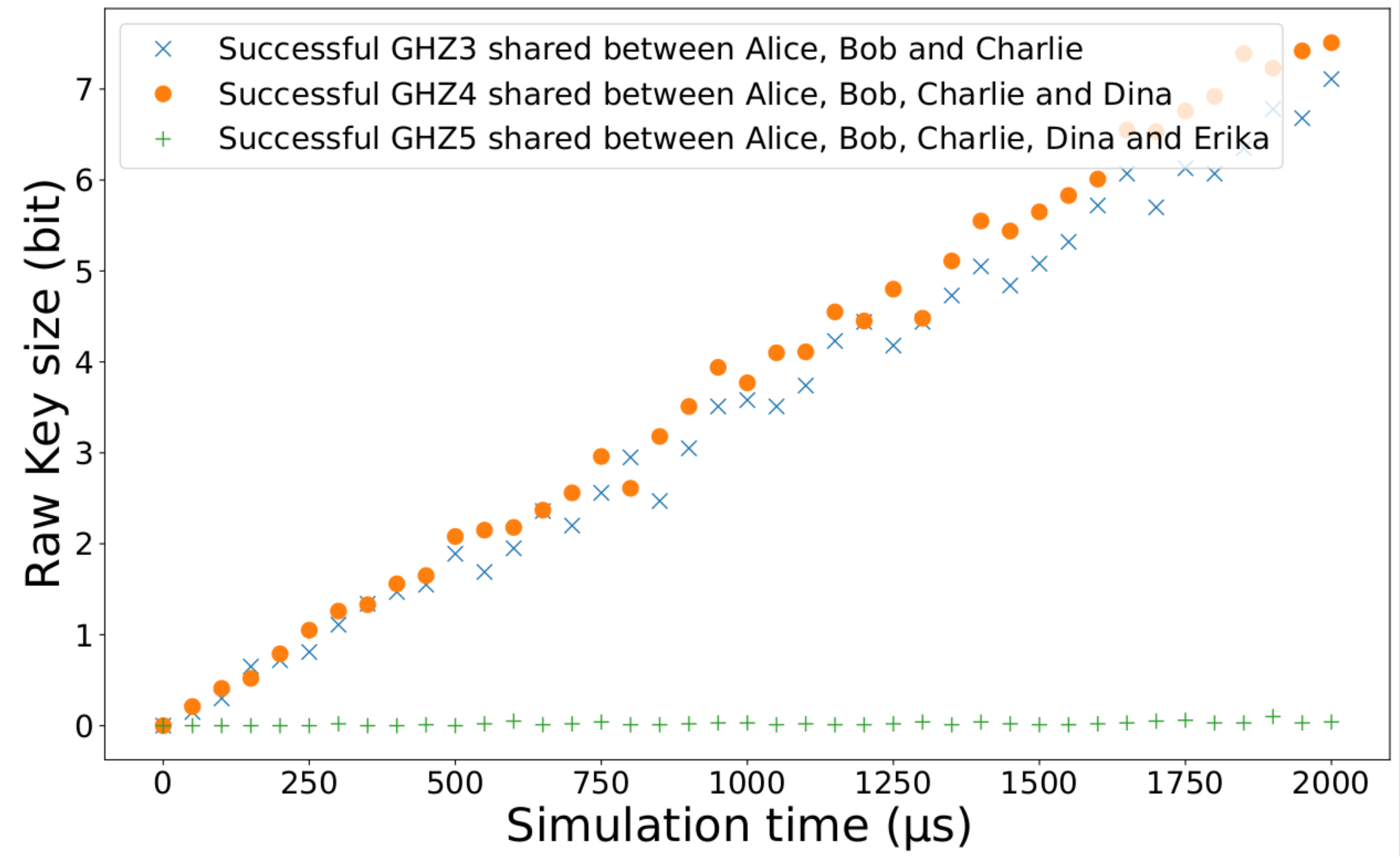}
    \caption{Number of GHZ states successfully transmitted to three, four and five Qlients as a function of the simulation time. It also corresponds to raw conference key size in a CKA context. Each point has been averaged over 100 runs of simulation.}
    \label{fig:GHZ}
\end{figure}

We also show in Fig.~\ref{fig:GHZ} the number of GHZ states that arrive successfully to 3, 4 or 5 Qlients as a function of the simulation time. This may correspond, for instance, to accumulated raw conference key, which can then be made secure following a conference key agreement protocol (see Sec.~\ref{sec:CKA}). We can see that scaling GHZ states to a larger size with the fusion operations considered in our simulations is challenging; going from  3 and 4 qubit GHZ states to 5 and 6 qubit GHZ states requires the operation to succeed twice, which occurs with low probability. Upgrading to techniques based on deterministic single-photon sources may offer a promising avenue towards such scaling~\cite{istrati2020}. \\

Once GHZ states are shared, multiparty quantum protocols such as conference key agreement or anonymous transmission become available to Qlients. 
When running, for instance, conference key agreement, the Qlients may not trust that the source is indeed providing GHZ states. The Qonnector may be dishonest or have noisy hardware. In this case, the Qlients will perform several verification rounds in between actual protocol rounds. Following the discussion in Sec.~\ref{sec:Verif}, it would take approximately 50~seconds for 4 Qlients to obtain one state such that there is a probability 0.99 that the state has 90\% fidelity with the GHZ state. By randomly doing key agreement rounds in between tens of thousands verification rounds, four Qlients thus can perform a conference key agreement protocol secure against a malicious Qonnector in a few tens of minutes. They would then get a raw key from which a common secret key can be extracted. Hence, in a Quantum City with currently available hardware, secure 4-party conference key agreement protocols can be achieved in a relatively practical amount of time.  \\

Similarly, the full anonymous message transmission protocol from \cite{Anonymity} supposes that the protocol presented Sec.~\ref{sec:Anonymous} is done in between GHZ verification rounds. This means a a message of a few bits can be securely and anonymously transmitted in less that an hour in our Paris Quantum City setting when 4 Qlients are involved. 
However, the time it would take to perform these protocols for more Qlients becomes impractical with present technology because of the low probability of $n$-qubit GHZ state generation when $n>4$ and of the high overhead in GHZ states required by the verification protocol. 
Nevertheless if the Qlients choose to trust their Qonnector, the rate at which the presented protocol can be performed becomes realistic. We can then conclude that near-term multiparty quantum protocols will most probably have to work in a trusted-node scenario.

\section{From Quantum Cities to a Quantum Europe}

In this work, we have presented and simulated a simple near-term network architecture that is realistic for metropolitan-scale deployment. It consists of a single powerful node with the ability to create bipartite or multipartite entanglement and make joint measurements to pairs of qubits, that we call the Qonnector, and a number of users with simple capabilities allowing them to prepare and measure single qubits, that we call the Qlients. Together they form a Quantum City. \\

\begin{figure}[!ht]
     \centering
     \includegraphics[width=13.5cm]{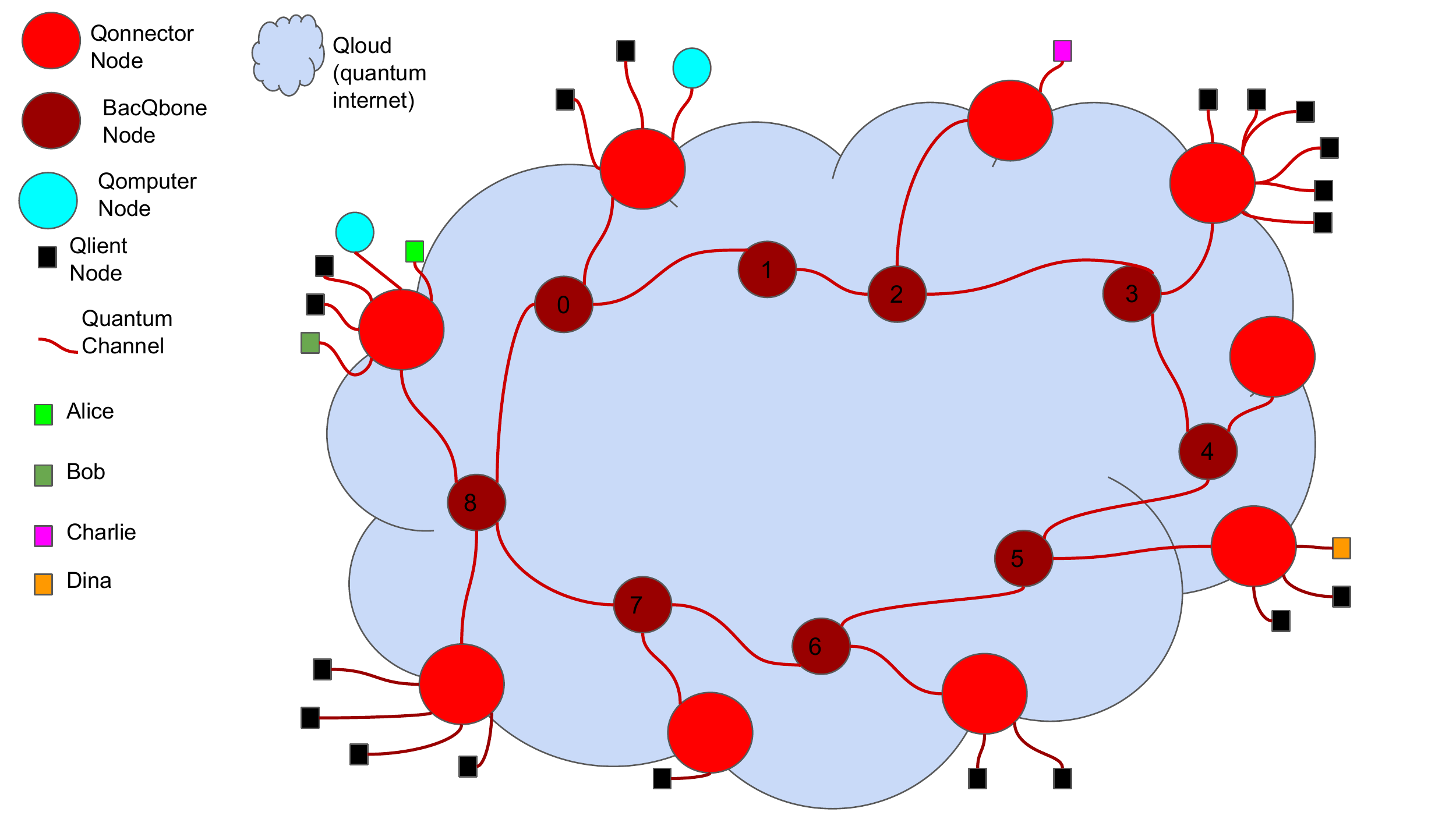}
     \caption{Schematic drawing of a Qloud: a possible way to connect Quantum Cities towards a global quantum network.}
     \label{fig:Qloud}
 \end{figure}

In the longer term, metropolitan quantum networks such as our Quantum City could be linked together using quantum repeaters or satellite links, to create a larger-scale quantum communication network, see Fig.~\ref{fig:Qloud}. In such a network, or Qloud, Qonnectors can be linked through BacQbone nodes materialized with quantum repeaters or satellites, and Qlients of different Quantum Cities can access the network through their nearest Qonnector. The underlying principle is to facilitate connectivity via centralized nodes and upgradability, both of which are key network aspects.

Our network simulation analysis of the performance of various realistic protocols pertinent to the Quantum City setting contributes to large ongoing efforts towards a global vision of the Quantum Internet. These include, for instance, the work of the IETF Quantum Internet Research Group \cite{QIRG}, an international collaboration of researchers whose goal is to define common grounds for research in quantum networks towards the Quantum Internet. The overarching objective is to construct architectural principles for flexible and scalable development of quantum networks, and according to these principles, enhance today's networking capabilities and applications.\\

To this end, it is important for the network to allow for growth and adaptability to tomorrow's applications to avoid changing significant parts of the hardware at each new protocol generation. Supporting hardware heterogeneity is also crucial as many new techniques are still under investigation. Ensuring security is an important asset, and the network has to be easy to manage and monitor and be resilient to failure and to malicious actors. Accessing to benefits of such a network as soon as possible is also highly desirable. Our simulations illustrate that such features are possible even with present or near-term technology.\\


Our work also contributes to the vision of the Quantum Internet Alliance~\cite{QIA}, a large European scientific collaboration, which has defined the expected stages of development for quantum networks~\cite{QIavision}, and pursues the study of all layers of a quantum version of the OSI network model, from hardware maturity to applications. While the protocols and simulations presented here are in the so-called prepare-and-measure and entanglement stages of quantum network development, rapid advances in user nodes and in the integration of quantum storage and repeating capabilities, could rapidly open new possibilities. In this context, the feasibility of actual quantum network deployment in Europe in the near future is currently under intense investigation. In a companion work, we will present realistic simulations of a Qloud in a European setting, including simulation of satellite-based links.
 
 \section{Conclusion and open questions}
 
In the network simulations we have presented, we have assumed limited hardware requirements for the network users, the Qlients, which are connected to one central and powerful node, the Qonnector. We have shown through different simulations how several applications can be made available to Qlients and have tried to give concrete estimations of their performance. The NetSquid code that we have used to produce our results is available on GitHub~\cite{github}. We have showcased our analysis on a specific network instance based on realistic hardware parameters and actual locations in the Paris region. \\

Our work gives rise to a number of open questions. Notably, we did not consider quantum memories in the nodes of our Quantum City. Integrating this aspect is of particular interest as it will allow for efficient routing strategies between the Qlients and for on-demand operations. Synchronization and timing strategies that need to be put in place at the nodes were also not discussed, and are crucial for proper network operation and hence for a complete protocol analysis. It will also be important to extend NetSquid to support the simulation of coherent state generation and coherent detection techniques to allow the investigation of an even wider range of protocols and applications. Finally, we also did not consider the effect of noise such as dark counts in Bell state measurements. Future work will also include more detailed error models for some of the protocol operations. \\

Our focus here has been to explore what applications can be available with optimized and realistic resources to quantum network users today. Our results highlight the significance and relevance of early deployment of quantum networks, while also preparing the ground for applications that will become available when more advanced quantum hardware is integrated, thereby unlocking the full potential of a Quantum Internet.


\section*{Acknowledgements}
We would like to thank Pierre-Emmanuel Emeriau, Robert Booth, Luka Music, Ulysse Chabaud, Léo Colisson, Paul Hilaire, Liao Chin-Te, Marie Billard and Frédéric Grosshans for fruitful discussions. This project is part of the Quantum Internet Alliance and has received funding from the European Union’s Horizon 2020 research and innovation program under grant agreement No 820445. We also acknowledge support from the European Union through the Project ERC-2017-STG-758911 QUSCO, from QuantERA through the project QuantAlgo and from the French National Research Agency (ANR) through the Project SoLuQS. 
\bibliographystyle{unsrt}
\bibliography{article}

\end{document}